\documentclass[]{JHEP3}
\usepackage{graphicx}
\JHEPspecialurl{http://jhep.sissa.it/JOURNAL/JHEP3.tar.gz}

\usepackage{amsmath}
\usepackage{epsfig,multicol,bbm}
\usepackage{url}

\newcommand{\newc}{\newcommand}
\newc{\gsim}{\lower.7ex\hbox{$\;\stackrel{\textstyle>}{\sim}\;$}}
\newc{\lsim}{\lower.7ex\hbox{$\;\stackrel{\textstyle<}{\sim}\;$}}
\newc{\gev}{\,{\rm GeV}}
\newc{\mev}{\,{\rm MeV}}
\newc{\ev}{\,{\rm eV}}
\newc{\kev}{\,{\rm keV}}
\newc{\tev}{\,{\rm TeV}}

\def\Im{\mathop{\rm Im}}

\newc{\mz}{M_Z}
\newc{\mpl}{M_*}
\newc{\mw}{m_{\rm weak}}
\newc{\nr}[1]{N^c_R{}_{#1}}
\usepackage{amsmath}
\def\beq{\begin{equation}}
\def\eeq{\end{equation}}
\def\bea{\begin{eqnarray}}
\def\eea{\end{eqnarray}}
\def\bitem{\begin{itemize}}
\def\eitem{\end{itemize}}
\newcommand{\bec}{\begin{center}}
\newcommand{\eec}{\end{center}}
\newc{\ie}{{\it i.e.}}          \newc{\etal}{{\it et al.}}
\newc{\eg}{{\it e.g.}}          \newc{\etc}{{\it etc.}}
\newc{\cf}{{\it c.f.}}

\def\bar#1{\overline{#1}}
\def\vev#1{\left\langle #1 \right\rangle}

\def\abs#1{\left| #1\right|}

\def\inv{^{\raise.15ex\hbox{${\scriptscriptstyle -}$}\kern-.05em 1}}
\def\lbar{{\lower.35ex\hbox{$\mathchar'26$}\mkern-10mu\lambda}} 

\def\to{\rightarrow}

\let\<=\langle
\let\>=\rangle

\let\+=\uparrow
\let\al=\alpha
\let\be=\beta
\let\ga=\gamma
\let\Ga=\Gamma
\let\de=\delta
\let\De=\Delta
\let\ep=\epsilon

\let\la=\lambda

\let\Om=\Omega
\newcommand\fverb{\setbox\fverbbox=\hbox\bgroup\verb}
\newcommand\fverbdo{\egroup\medskip\noindent%
			\fbox{\unhbox\fverbbox}\ }
\newcommand\fverbit{\egroup\item[\fbox{\unhbox\fverbbox}]}
\newbox\fverbbox

\title{A Unified Theory of Matter Genesis: \\
Asymmetric Freeze-In}

\date{\today}

\author{Lawrence J. Hall\\
	Department of Physics, University of California, Berkeley and\\
Theoretical Physics Group, LBNL, Berkeley, CA 94720, USA, {\rm and}\\
Institute for the Physics and Mathematics of the Universe, \\ University of Tokyo, Kashiwa 277-8568, Japan\\
	E-mail: \email{ljhall@lbl.gov}}

\author{John March-Russell\\
	Rudolf Peierls Centre for Theoretical Physics, University of Oxford, 1 Keble Road, Oxford, OX1 3NP, UK\\
	E-mail: \email{jmr@thphys.ox.ac.uk}}

\author{Stephen M. West\\
	Royal Holloway, University of London, Egham, TW20 0EX, UK, {\rm and}\\
	Rutherford Appleton Laboratory, Chilton, Didcot, OX11 0QX, UK\\
	E-mail: \email{Stephen.West@rhul.ac.uk}}

\preprint{OUTP-10-20P}	

\abstract{We propose a unified theory of dark matter (DM) genesis and baryogenesis.   It explains the observed link
between the DM density  and the baryon density, and is fully testable by a combination of collider experiments
and precision tests.    Our theory utilises the ``thermal freeze-in" mechanism of DM production, generating particle anti-particle   
asymmetries in decays from visible to hidden sectors.
Calculable, linked, asymmetries in baryon number and DM number are produced by the feeble
interaction mediating between the two sectors, while the out-of-equilibrium condition necessary 
for baryogenesis is provided by the different temperatures of the visible and hidden sectors.
An illustrative model is presented where the visible sector is the MSSM, with the relevant CP violation arising from phases in the gaugino and Higgsino masses, and both asymmetries are generated at temperatures of order 100 GeV.    
Experimental signals of this mechanism can be spectacular, including: long-lived metastable states late decaying at the LHC; apparent baryon-number or lepton-number violating signatures associated with these highly displaced vertices; EDM signals
correlated with the observed decay lifetimes and within reach of planned experiments; and a prediction for the mass of the dark matter particle that is sensitive to the spectrum of the visible sector and the nature of the electroweak phase transition.}

\keywords{Beyond Standard Model, Dark Matter, Baryogenesis}

\begin{document} 

\maketitle
\section{Introduction \label{intro}}

Traditionally, the processes of dark matter (DM) genesis and baryogenesis have been assumed to be independent, with the
DM density being determined by the mechanism of ``thermal freeze out" \cite{freezeout} leaving behind a relic density of DM
particles with no net conserved quantum number, while the baryon density is entirely determined by an asymmetry
generated by CP-violating and baryon-number-violating out-of-equilibrium processes.    A consequence of such decoupled
genesis mechanisms is that the DM-to-baryon ratio $\Om_{d}/\Om_b$ could {\it a priori} lie, with reasonable assumptions,
anywhere in the range $10^{10} - 10^{-10}$ in discord with the close coincidence observed, $\Om_{d}/\Om_b
\simeq 4.86$ \cite{rpp}.

In this work we will argue that there exists an elegant and fully calculable class of theories in which the baryon asymmetry and DM density are directly linked, explaining the observed coincidence.\footnote{In this paper we investigate the possibility that $\Om_b$ 
and $\Om_{d}$ are dynamically linked.  Previous investigations along this line include \cite{others}.   An alternate
possibility is that the ratio $\Om_{d}/\Om_b$ is anthropically determined by environmental selection on gross features, as for axion dark matter in the regime of large axion decay constant, $f_{PQ} > 10^{12} \gev$ \cite{axionmiss}.  One difficulty with the anthropic approach is that the observed ratio is not close to the critical boundary
$\Om_{d}/\Om_b \sim 200$ derived from gross requirements such as successful structure formation \cite{axionmiss,anthropic}.}  The mechanism simultaneously generates both baryon and DM asymmetries, and does not rely on any pre-existing asymmetry from high temperatures.   Moreover, it does not use  ``thermal freeze-out", but rather a variant of the recently suggested mechanism of ``thermal freeze-in" \cite{FIMP}\footnote{For related discussions of the production of the RHD sneutrino and neutrino see also \cite{rhneut}.}, which, in advantageous cases, is IR dominated by low temperatures 
and therefore independent of the uncertain early thermal history of the universe and possible new interactions at high scales.  The relic abundances reflect a combination of initial thermal distributions together with particle masses and couplings that can be measured in the laboratory, allowing, in advantageous cases, the entire mechanism of baryogenesis and DM-genesis to be experimentally explored and confirmed. 

\begin{figure}
\vspace{-2cm}
\centerline{\includegraphics[width=12cm]{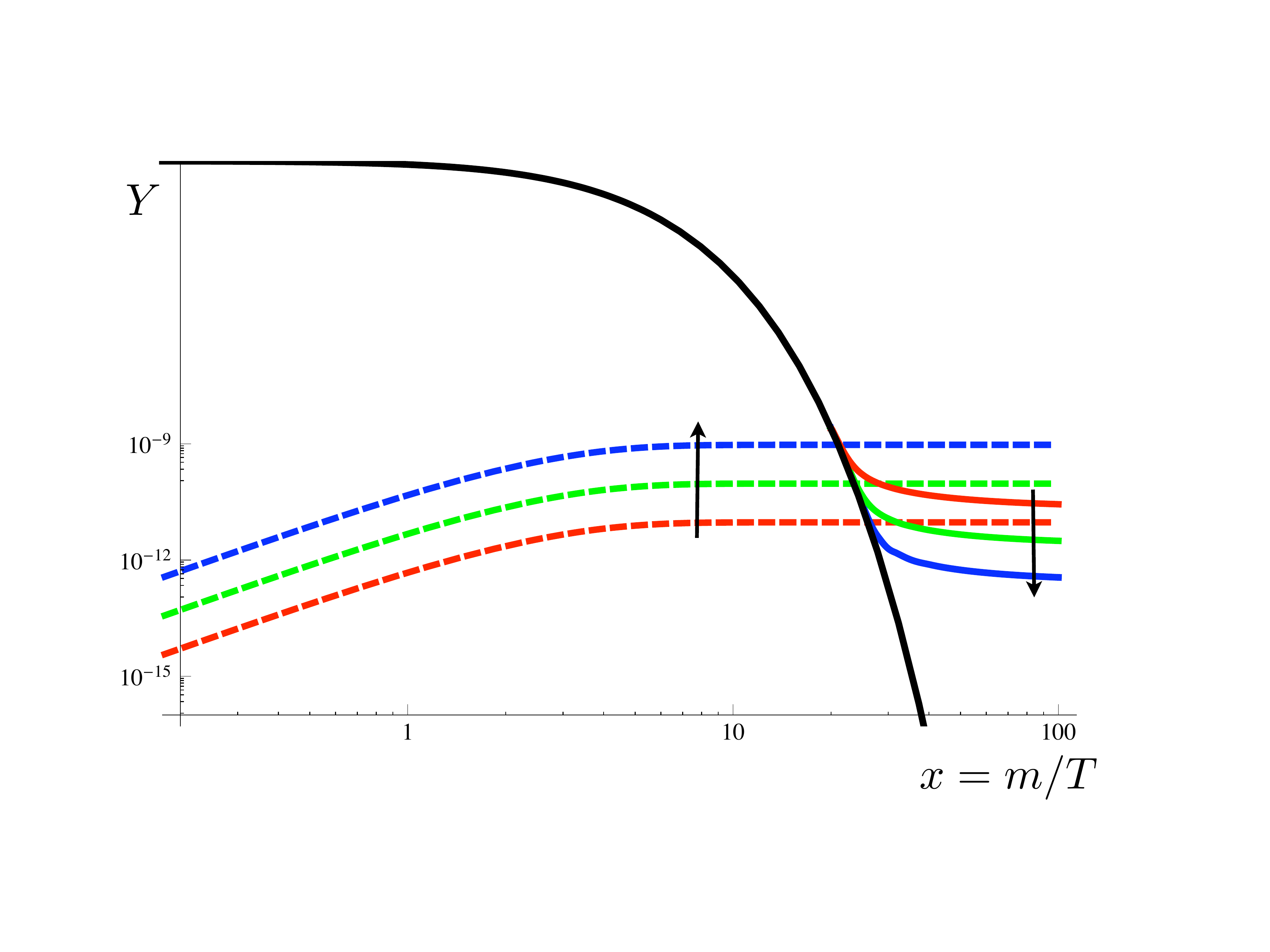}}
\vspace{-1.5cm}
\caption{From Ref.\cite{FIMP}.  Log-Log plot of the evolution of the relic yields for freeze-in via a Yukawa interaction (dashed coloured) and conventional freeze-out (solid coloured) as a function of $x=m/T$.  Arrows indicate the effect of increasing coupling strength for the two processes, and the black solid line indicates yield if equilibrium is maintained. The freeze-in yield is dominated by the epoch $x\sim 2-5$, in contrast to freeze-out which only departs from equilibrium for $x\sim 20-30$.\label{fi_vs_fo}}
\end{figure}

Since our genesis mechanism is based on thermal freeze-in, it behooves us to start with a description of this process.  Suppose there is a set of bath particles that are in thermal equilibrium at temperature $T$ and some other particle, $X$, having couplings with the bath that are so feeble that $X$ is thermally decoupled from the plasma.   Although feeble, the interactions with the bath do lead to some $X$ production and, for renormalizable interactions, the dominant production of $X$ occurs as $T$ drops below the mass of the lightest bath particle
coupling to $X$ (providing $X$ is lighter than the bath particles with which it interacts). The abundance of $X$ ``freezes-in" with a yield that {\it increases} with the interaction strength of $X$ with the bath, in contradistinction to traditional freeze-out which begins with a full $T^3$ thermal number density of DM particles, and where {\it reducing} the interaction strength helps to maintain this large abundance.  As the temperature drops below the mass of the relevant particle, the DM is either heading away from (freeze-out) or towards (freeze-in) thermal equilibrium.  These trends are illustrated in Figure~\ref{fi_vs_fo} (from Ref.\cite{FIMP}) which shows the evolution with temperature of the DM abundance
according to freeze-in and conventional freeze-out.   Freeze-in provides the only possible alternative thermal production mechanism that is dominated by IR processes, and so can, in principle, be completely tested and confirmed at colliders without
knowledge of UV interactions and the complete thermal history of the early universe, in particular the post-inflationary
reheat temperature $T_R$ (as long as it is above the weak scale).   

The freeze-in density is dominated, where possible, by decays or inverse decays involving the bath particles and $X$.  This is typically the case when some subset of the bath particles also carry the conserved quantum number that stabilizes $X$ (here we assume that the symmetry is a $Z_2$ parity).  Let $B$ be a $Z_2$-odd particle in the bath with mass $m_B$ that decays to $X$ with a small decay rate $\Gamma$.   As $T$ drops below $m_B$ and $B$ becomes non-relativistic the freeze-in process gives an $X$ yield
\beq
Y_X = \frac{n_X}{s} = C_{FI} \, \frac{M_{Pl} \Gamma}{m_B^2}
\label{FIyield}
\eeq
where $C_{FI} = 1.64 \, g_X/g^S_* \sqrt{g^{\rho}_*}$, corresponding to an $X$ relic-abundance 
\beq
\Om_X h^2=\frac{1.09\times 10^{27}}{g^S_* \sqrt{g^{\rho}_*}} \, \frac{m_X \Ga}{m_B^2}
\eeq
(see eq.~(6.10) of Ref.\cite{FIMP} and surrounding text for a fuller discussion).   This expression applies to the case in which there is no asymmetry in quantum numbers, and the $X$ states do not annihilate away to lighter states via interactions in the X-sector.\footnote{For
renormalizable interactions the yield, $Y_X$, is dominated by low temperatures due to the increase in both the Hubble doubling time, $1/H \sim M_{Pl}/T^2$, and production cross sections as the temperature $T$ lowers towards the relevant particle masses.  For non-renormalizable
interactions the FI yield is still calculable from thermal distributions, but depends on the unknown reheat temperature, $T_R$ \cite{FIMP}.}

Now suppose we have two sectors\footnote{For a general discussion of freeze-in and freeze-out processes in two sector cosmologies see \cite{twosectorcos}.}: the visible sector containing $B$ at temperature $T$ and a hidden sector containing $X$ at temperature $T'<T$.  (This initial condition can result from preferential inflaton decay to the SM sector.)  In the limit that there are no interactions connecting the sectors the lightest $Z_2$ odd particle of each sector will be stable.   We take the visible sector interactions to conserve $B-L$ and the hidden sector interactions to conserve a global $U(1)_X$, under which $X$ is charged \footnote{At some level we expect all continuous global symmetries to be violated, but here we are taking
$U(1)_X$ to be violated no more strongly than $B$-number is by non-perturbative effects in the SM, or by non-renormalizable terms suppressed by some high scale such as $M_{GUT}$ or $M_{pl}$ as in extensions of the SM.}.  We take an initial condition with no particle anti-particle asymmetries: $\eta_{B-L} = \eta_X = 0$.   Next we introduce an interaction $\De{\cal L} = \la {\cal O}$ that connects the two sectors, where the operator ${\cal O}$ transforms under both $B-L$ and $U(1)_X$, but preserves the combination $B-L+X$.
It is crucial that $\la$ is small enough that this interaction does not equilibrate the two sectors to a single temperature during the weak era.  Including powers of $m_B$ in ${\cal O}$ to make $\la$ dimensionless, this implies that $\la < 10^{-6} \sqrt{m_B/100 \gev}$.

\begin{figure}
\vspace{-2cm}
\centerline{\includegraphics[width=14cm]{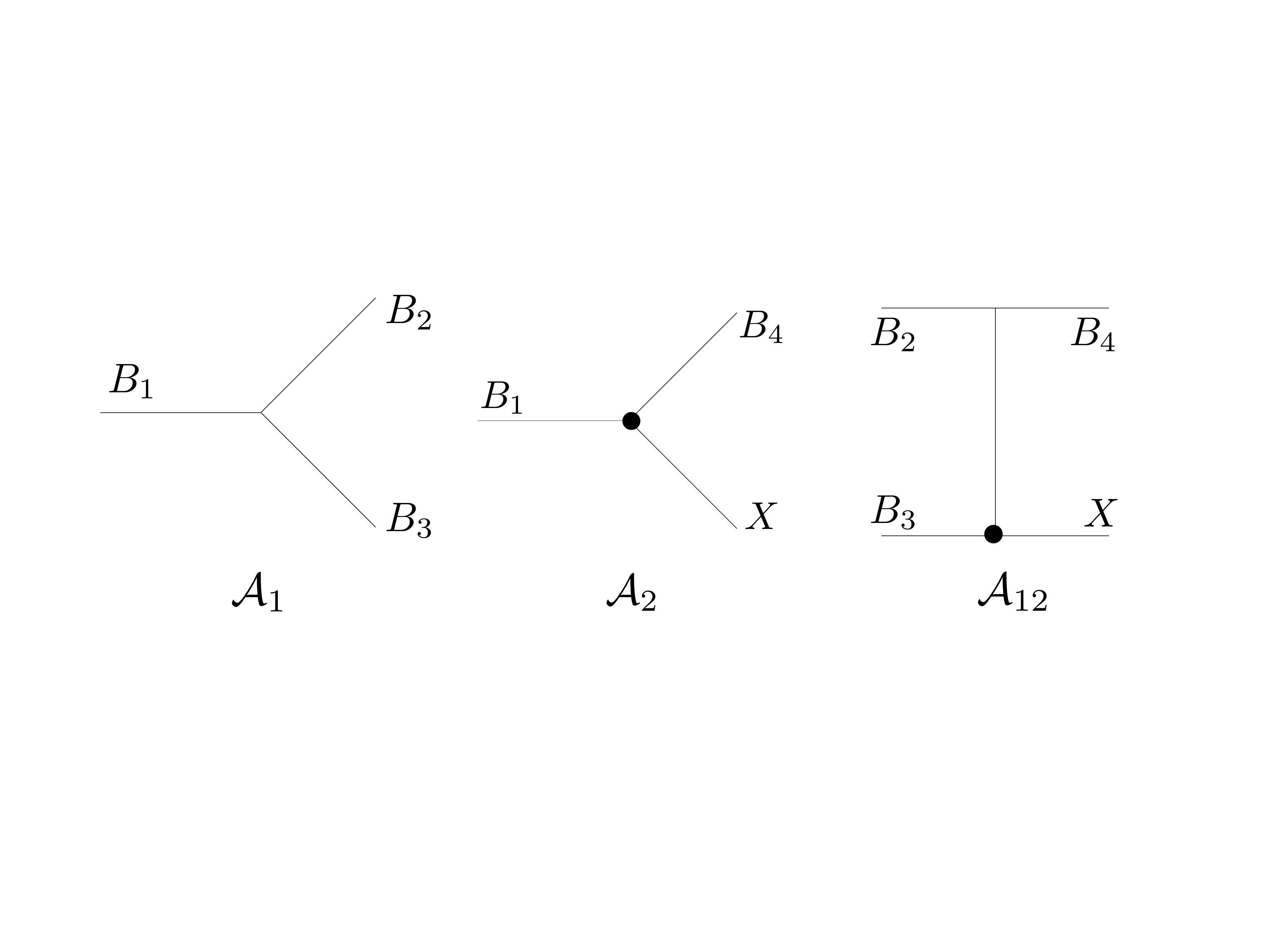}}
\vspace{-2.5cm}
\caption{Schematic illustration of the three on-shell amplitudes necessary for generation of the linked $(B-L)$ and $X$ asymmetries.  State
$B_1$ of the visible thermal bath must be able to decay to two different final states $B_2B_3$ and $B_4X$ with differing $(B-L)$ and $X$ quantum numbers, with respective amplitudes ${\cal A}_1$ and ${\cal A}_2$.  Moreover there must be an independent re-scattering process linking these differing final states, with amplitude denoted ${\cal A}_{12}$.   The black dot indicates the feeble coupling linking the visible and hidden sectors, while all other vertices are ${\cal O}(1)$ visible sector couplings.  The decay amplitude ${\cal A}_1$ implies that $B_1$ cannot be the lightest visible sector $Z_2$-odd particle.}
\label{Bdecayfull}
\end{figure}

Starting with equal densities of $B$ and its anti-particle $\bar{B}$, can decays of $B \rightarrow X + ...$ and $\bar{B} \rightarrow \bar{X} + ...$ generate an asymmetry $\eta_X$? If so then conservation of $B-L+X$ will ensure that at the same time an equal and opposite $B-L$ asymmetry is generated
\beq
\eta_{B-L} = -\eta_X \neq 0 .
\eeq
To accomplish this $B$ must have at least two different decay modes with different $X$ charges of the final state.  Consider, as schematically illustrated in Figure~\ref{Bdecayfull}, the simple case that the leading decay mode of $B$ occurs entirely within the visible sector with amplitude ${\cal A}_1$ and that the sub-leading decay to $X$ has amplitude ${\cal A}_2$.  An asymmetry $\eta_X$ is generated at 1-loop if there is re-scattering between these two final states, which we denote by the amplitude ${\cal A}_{12}$, and the asymmetry is given by
\beq
\eta_X = \epsilon Y_X \simeq C_{FI} \frac{M_{Pl}}{m_B} \, \frac{Im ({\cal A}_1 {\cal A}_{12} {\cal A}_2^* )}{128 \pi^3}
 \label{eq:etaX}
\eeq
where ${\cal A}_{1,2,12}$ have been made dimensionless by introducing factors of $m_B$.  Here $\epsilon$ is the asymmetry produced per decay of a $B \bar{B}$ pair to final states containing an $X$ state.  The novel feature of this genesis mechanism is the way in which the out-of-thermal equilibrium constraint is met. All initial particles involved in the reaction (and the inverse reaction) have thermal distributions.  However, the thermal distributions of $B$ and $X$ reflect {\it different} temperatures, so that there is no equilibration between the reaction and inverse reaction.   To get a feel for the likely magnitude of this asymmetry, we note that while $|{\cal A}_1|$ can be order unity, $|{\cal A}_2|$ and $|{\cal A}_{12}|$, which connect visible and hidden sectors, must both be less than $10^{-6} \sqrt{100 \gev/m_B}$, to avoid equilibration of the two sectors.  Hence we write
\beq
 \eta_X \simeq 10^{-2} \, \sin \phi \; \frac{C_{FI}}{10^{-3}}  \,  |{\cal A}_1| \,
 \left( \frac{| {\cal A}_2|}{10^{-6} \sqrt{m_B/100 \gev}}\right) \, \left(\frac{| {\cal A}_{12}|}{10^{-6} \sqrt{m_B/100 \gev}}\right) ,
 \label{eq:etaX2}
\eeq
where $\phi$ is the phase of ${\cal A}_1 {\cal A}_2^* {\cal A}_{12}$.   Hence we conclude that asymmetric freeze-in is easily able to generate a sufficiently large asymmetry.  The maximal asymmetry of $10^{-2}$ arises from the $g_*$ factors in $C_{FI}$ and the combinatorial factor of $128 \pi^3$.  The limit of $10^{-6} \sqrt{m_B/100 \gev}$ on the amplitudes that connect the two sectors arises from $\sqrt{m_B/M_{Pl}}$, and these cancel the factor of $M_{Pl}/m_B$ in eq.(\ref{eq:etaX}).

We emphasize that our idea is quite different and more ambitious than almost all previous attempts to link the baryon number
and DM densities \cite{others}.  The interaction $\la {\cal O}$  does not just transfer a pre-existing asymmetry in one sector, generated by some other earlier and unspecified mechanism, to the second sector: rather {\it this interaction simultaneously causes the
asymmetries in both sectors}.   Moreover, will show that this feature implies that the entire mechanism of simultaneous
baryogenesis and DM genesis can be explored by a combination of present and future collider experiments, and precision
measurements of CP-violating observables such as EDMs, again in contrast to most previously suggested mechanisms.  

For this idea to explain the baryon-to-DM ratio, however, some further challenges must be met:

\bitem

\item
If the only symmetry of consequence is $(B-L+X)$, then either the DM particle can decay to baryons 
or vice versa.  Stabilization of both baryons and DM requires a new symmetry, e.g., $R$-parity in the supersymmetric
(SUSY) case or a similar discrete symmetry in the non-supersymmetric case.

\item  The communication between the two sectors also generates a symmetric part of $X$ which in general dominates over the
$X$ asymmetry by a large factor $1/\ep$, where $\ep\lsim 10^{-3}$ is related to the (necessarily loop-suppressed) amount of physical CP-violation in the interactions.  We must efficiently annihilate this symmetric part of $X$ if the final DM density is to be determined
by the $X$ asymmetry and we are to successfully link the baryon density to the DM density.   Thus the second sector can not just be a
single feebly-interacting particle, but must be a {\it sector} with sizeable couplings and access to light states, either in the sector
itself, or in the SM, to which the symmetric part of $X$ can annihilate.\footnote{Since, by the hidden-sector non-thermalization
condition, all couplings to the SM must be feeble, annihilation of the symmetric part to SM states must be a two stage process 
ending with a non-symmetry-protected hidden sector state late decaying to light SM states via a feeble coupling. 
This is discussed in detail in Section~\ref{Xsector}.}

\item
Even in the case in which the symmetric part of the $X$ density is annihilated away, and we are left with just the asymmetry,
the baryon-to-DM density depends on the mass ratio $m_p/m_X$ as well as $\eta_b / \eta_X$.   Thus although, in the class of theories
we will discuss, the $\eta_b$ and  $\eta_X$ are linked (with some mild dependence on the weak-scale visible spectrum via details
of sphaleron-mediated equilibration), unless we have a dynamical reason why $m_p/m_X \sim 1$ we have not really explained
the baryon-to-DM density.   We will argue that this problem is naturally solved in supersymmetric theories with
supergravity or anomaly mediation of supersymmetry breaking.   

\eitem

If, in addition, we impose the additional requirement that this mechanism of DM-genesis be fully testable at colliders, then we must be able to measure $m_X$, requiring in turn that we be able to see it via, eg., lightest-ordinary-sector-superparticle (LOSP) decay.  This implies that although the hidden sector could possess many states charged under $U(1)_X$, the stable state is the one that appears in ${\cal O}$.  In the
following we will always impose this extra condition, though we emphasize that this is not necessary for the success of the genesis mechanism itself, only for its complete testability at colliders and in EDM experiments.

\section{Minimal Supersymmetric Models \label{model}}

Consider a supersymmetric extension of the SM, which we here take to be the MSSM appended by an R-parity odd, SM-gauge-singlet chiral superfield, $X$, as well as some $U(1)_X$-preserving $X$-sector interactions to be discussed in Section~\ref{Xsector}.  All of the operators of lowest dimension, 4 and 5, that couple $X$ to the MSSM sector are
\beq
L_i H_u X ; \qquad  L_i L_j {\bar E}_k X, \qquad L_i Q_j {\bar D}_k X, \qquad {\bar U}_i {\bar D}_j {\bar D}_kX, \qquad L_i H_d^\dagger (X, X^\dagger ), \qquad L_i H_u X^\dagger
\label{rpodd}
\eeq
where $i,j,k$ are generation indices and it is understood that (non-) holomorphic operators are (D) F terms.  The first operator is of dimension 4 and the rest are dimension 5.   If the hidden-sector field $X$ couples to ${\bar U}{\bar D}{\bar D}$ and at least one of the other operators, then, for the sizes of couplings and hidden-sector masses that will be of interest to us, too fast proton decay $p\to \pi^0 \ell^+, \pi^+ {\bar\nu}$ generally results. Therefore in the following we will always exclude this case.

For simplicity we will also assume in the following that the initial temperature of the hidden sector $T'\ll T_{SM}$, so that inverse decays
and scattering of hidden sector states back to the visible sector can be ignored.  It is straightforward to relax this assumption leading to
only small corrections to our formulae as long as $T'$ remains substantially smaller than $T_{SM}$ during the epoch of freeze-in.

\subsection{Asymmetric freeze-in via $LH_uX$} \label{LHX}

The most attractive case involves purely
\beq
\De W = \la_i L_i H_u X ,
\label{LHint}
\eeq
since, uniquely, the coupling to $X$ is renormalizable and the freeze-in yield is automatically IR dominated \cite{FIMP} and insensitive to the unknown high temperature history of the universe, such as the post inflationary reheat temperature, $T_R$ (as long as $T_R$ is above the weak scale).
Before non-perturbative SM effects are considered, the total theory is invariant under $U(1)_{(L-X)}$, linking total lepton number to $X$-number.  In the presence of anomalies and non-perturbative SM effects this gets modified to $U(1)_{(B-L+X)}$, allowing the transference
of linked $X$- and $L$ asymmetries to a baryon asymmetry.   The precise equilibration between the $L$ and $B$ asymmetries, and thus the link between $X$ and $B$ asymmetries and associated prediction for the DM mass, depends upon the details of the weak scale spectrum as discussed in Section~\ref{massprediction}.

The weak scale spectrum also determines which processes are dominantly responsible for generating the asymmetry.  For concreteness, consider the case where asymmetric freeze-in production of DM is dominantly due to decays of charginos or neutralinos.  Contributions from slepton decays may be suppressed kinematically or by small $A$ terms.  Since the generated $X$ and $(B-L)$ asymmetries are naturally small, cf eq.(\ref{eq:etaX2}), we also require new sources of CP-violation not suppressed by small Yukawas.  We therefore use the CP-violating phases of the gaugino-higgsino sectors that arise in the MSSM.  

The heavier charginos and neutralinos can decay either to final states with $L=X=0$ via standard MSSM interactions involving
$W$'s or $Z$'s, or they can decay to $\ell \phi_X$ states with $L=-X=\pm 1$ via the $\la LHX$ interaction.  
On the other hand, the LOSP state, be it a chargino or a neutralino, can only decay via the $\la LHX$ interaction, leading to a characteristic signature of our scenario, namely displaced lepton number violating vertices (in the case of the
other $R_p$-odd operators of eq.(\ref{rpodd}) displaced LOSP decays violating lepton or baryon number also occur,
see Section~\ref{BL-LHC}).

The heavier charginos or neutralinos possess simultaneously CP-violating and L/X-number-violating decays that lead to the generation of the asymmetry.   The CP-violating net $L$-number produced per chargino decay to $\ell \phi_X$ final states has the form
\beq
\ep_a^- =\sum_{k} \frac{\Ga(\tilde{\chi}_a^-\rightarrow l_k^-\phi_X)-\Ga(\tilde{\chi}_a^+\rightarrow l_k^+\phi^*_X)}{\Ga(\tilde{\chi}_a^-\rightarrow l_k^-\phi_X)+\Ga(\tilde{\chi}_a^+\rightarrow l_k^+\phi^*_X)}
\eeq
and similarly for neutralino decay
\beq
\ep_a^0 = \sum_{k}\frac{\Ga(\tilde{\chi}_a^0\rightarrow \nu_k\phi_X)-\Ga(\tilde{\chi}_a^0\rightarrow \bar{\nu}_k\phi^*_X)}{\Ga(\tilde{\chi}_a^0\rightarrow \nu_k\phi_X)+\Ga(\tilde{\chi}_a^0\rightarrow \bar{\nu}_k\phi^*_X)}
\eeq
where the sum over $k$ counts all possible final state flavours of lepton.  By $U(1)_{(L-X)}$ symmetry the net $X$-number produced per chargino and neutralino decay is just the opposite of these. Note that, unconventionally, we have 
not normalized these asymmetries by the total decay width of the initial states, but rather by the partial decay width to $\ell \phi_X$
states.   This allows us to express the asymmetry yield in the convenient form, $\ep Y_X$, as in eq.(\ref{eq:etaX}).
Expanding the matrix elements of the partial decay widths into tree and 1-loop terms as
\bea\nonumber
\mathcal{M}=C_0A_0+C_1A_1+...,
\eea
where the factors labelled by $C_i$ are collections of coupling constants and the $A_i$ are the associated tree and 1-loop amplitudes, we can write each individual asymmetry in the form (see eg, \cite{Davidson:2008bu} for a similar analysis in the leptogenesis case)
\beq
\ep=\sum_{n}\frac{2\Im{[C_{0n}C_{1n}^*]}\int \;d\Pi_{l,\phi}\tilde{\de}\Im{[A_{0n}A_{1n}^*]}}{\abs{C_{0n}}^2\int d\Pi_{l,\phi}\tilde{\de}\abs{A_{0n}}^2},
\label{cutformula}
\eeq
where 
\beq\nonumber
d\Pi_{l,\phi}=d\Pi_{l}d\Pi_{\phi}=\frac{d^3p_l}{2E_l(2\pi)^3}\;\frac{d^3p_{\phi_X}}{2E_{\phi_X}(2\pi)^3},\qquad \tilde{\de}=(2\pi)^4\de^4(P_{i}-P_f)
\eeq
and $P_i$, $P_f$ are the incoming and outgoing four-momentum ($P_f=P_l+P_{\phi_X}$).  The sum over $n$ represents all the possible diagrams that may contribute to the asymmetry. As usual, the loop amplitude $A_1$ only has an imaginary part when there are branch cuts due to intermediate on-shell particles.  The position of the cut in the diagrams is also crucial: In order to generate an asymmetry in $L$ number the value of $L$ at the cut must be different compared to that of the final states \cite{Kolb:1979qa}.  
This statement has important consequences for which diagrams will lead to an asymmetry.  In particular, LOSP decays cannot lead to the generation of a $X$ asymmetry: By definition, there is no MSSM $R_p$-odd state lighter than the LOSP which can be on shell in the loop,  while $\phi_X$, which is $R_p$ odd and lighter than the LOSP, and thus could appear on-shell in the loop, does not satisfy the $L$- (or $X$-) number cut condition.  On the other hand, the decays of heavier MSSM states, such as the heavier neutralinos and charginos, can have intermediate on-shell states which satisfy the necessary conditions.

\begin{figure}[tb] 
\vspace{-10mm}
$\begin{array}{cc}
\hspace{-10mm}
\includegraphics[width=0.37\textwidth]{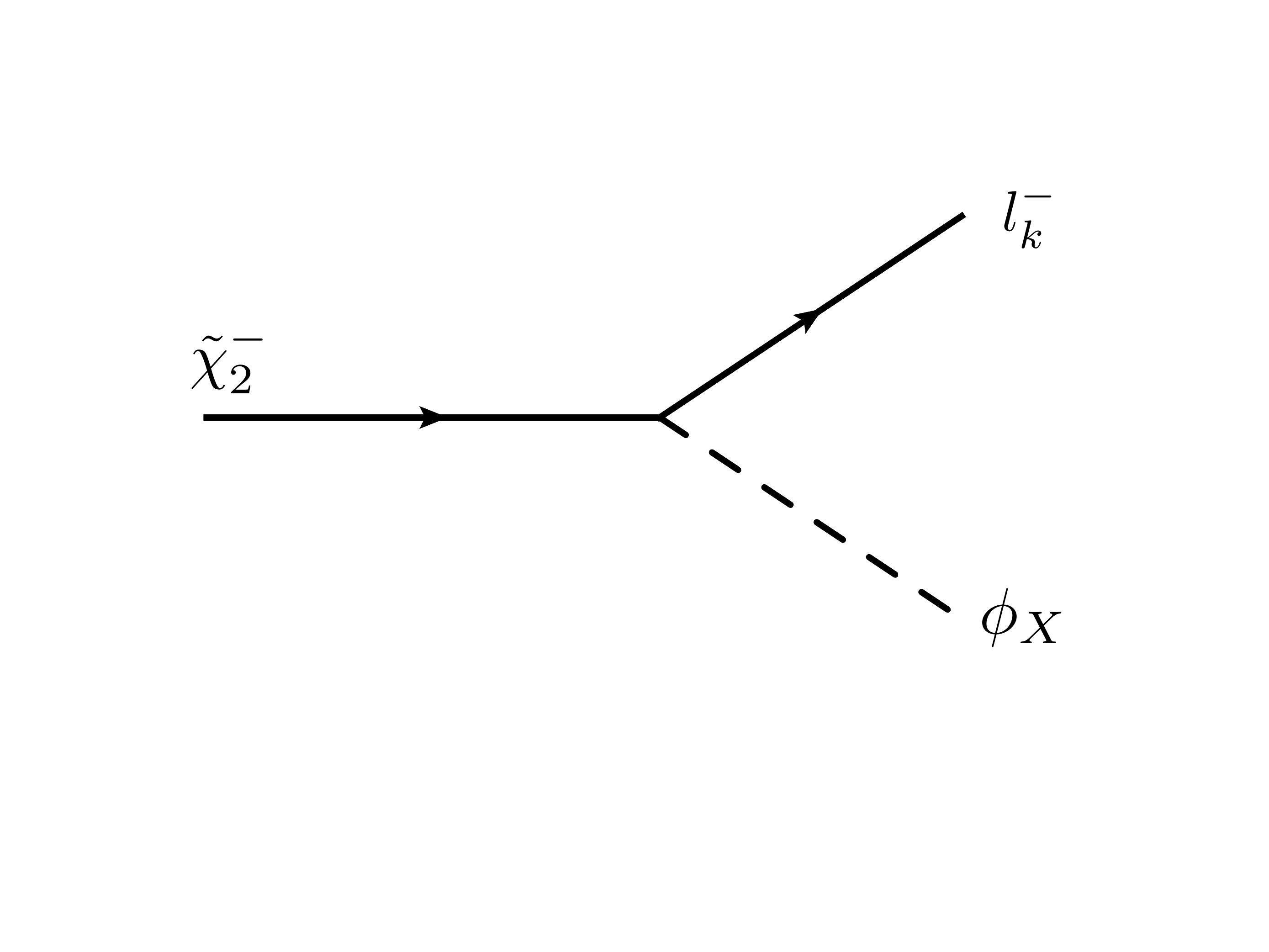}\hspace{-9mm}
\includegraphics[width=0.37\textwidth]{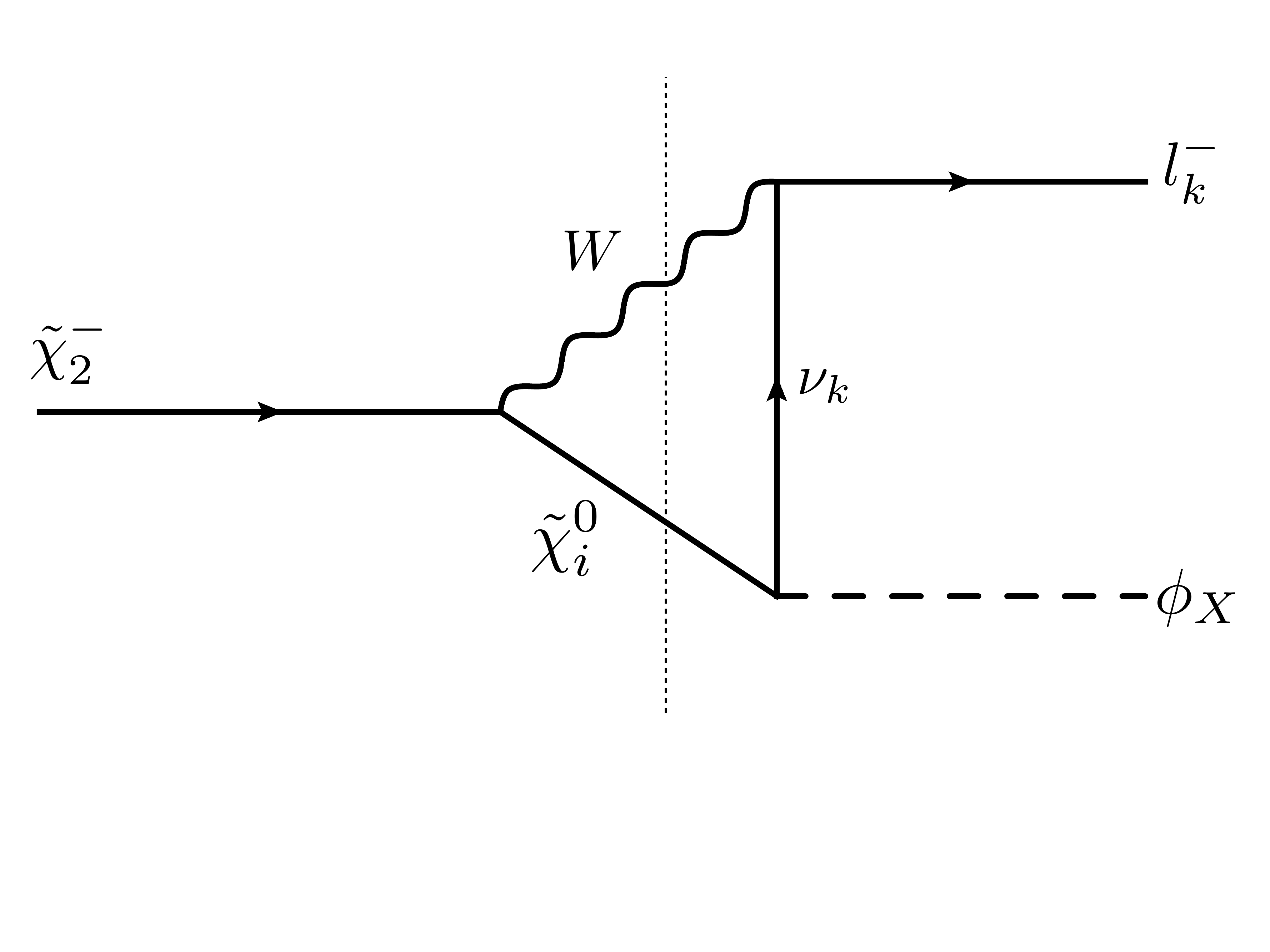}\hspace{-1mm}
\includegraphics[width=0.40\textwidth]{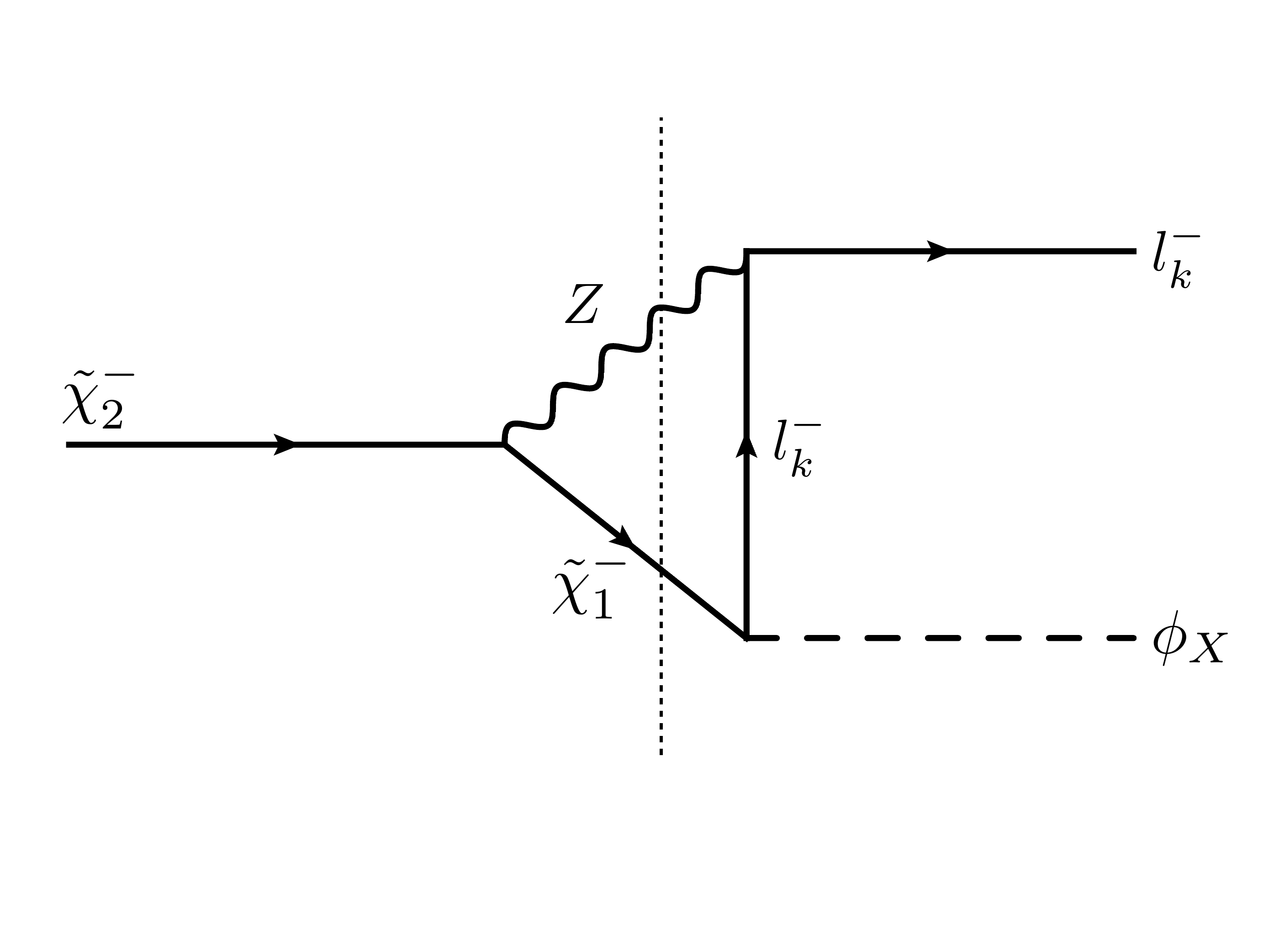} \\\hspace{-2mm}
\includegraphics[width=0.40\textwidth]{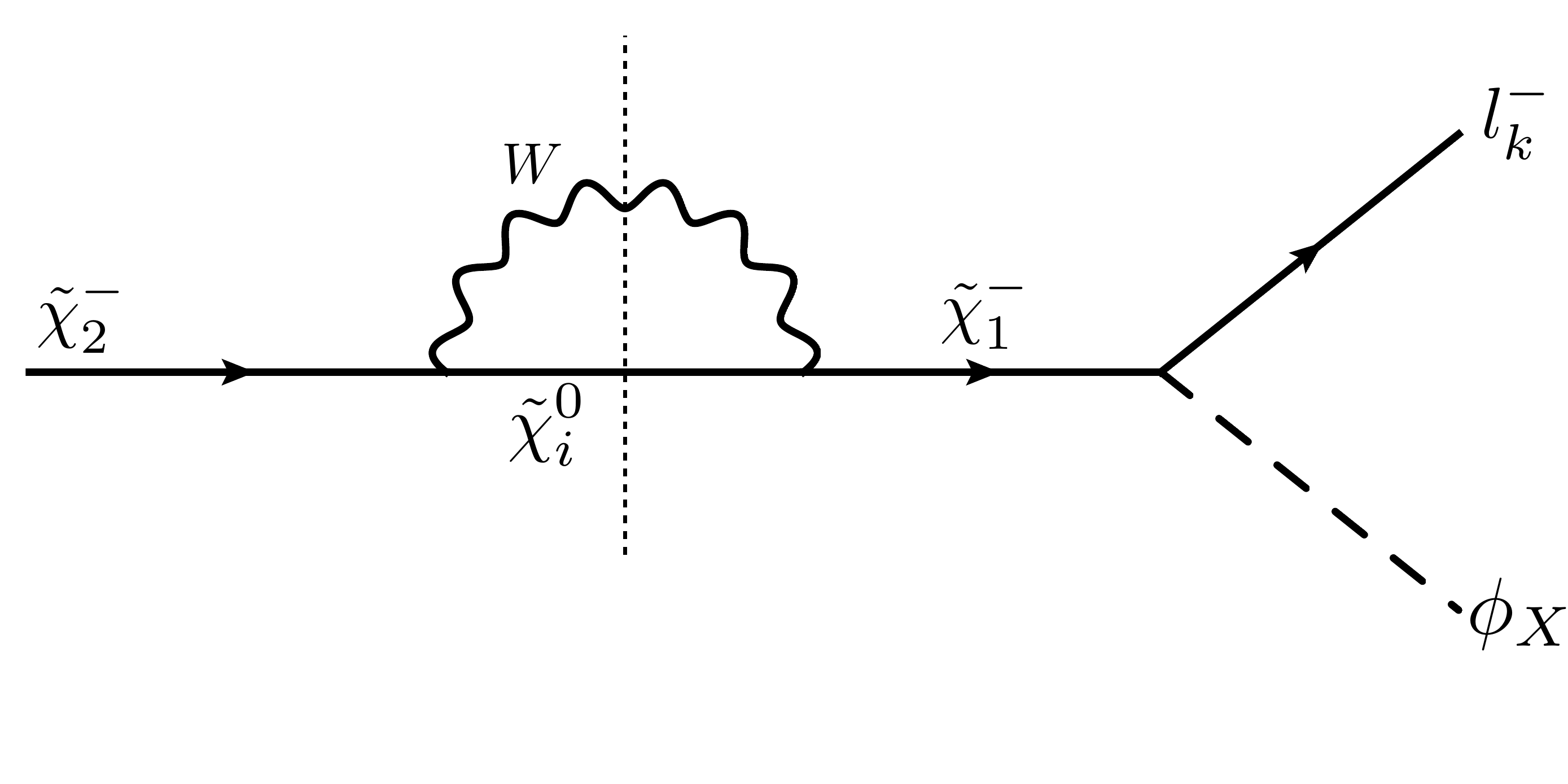}
\includegraphics[width=0.40\textwidth]{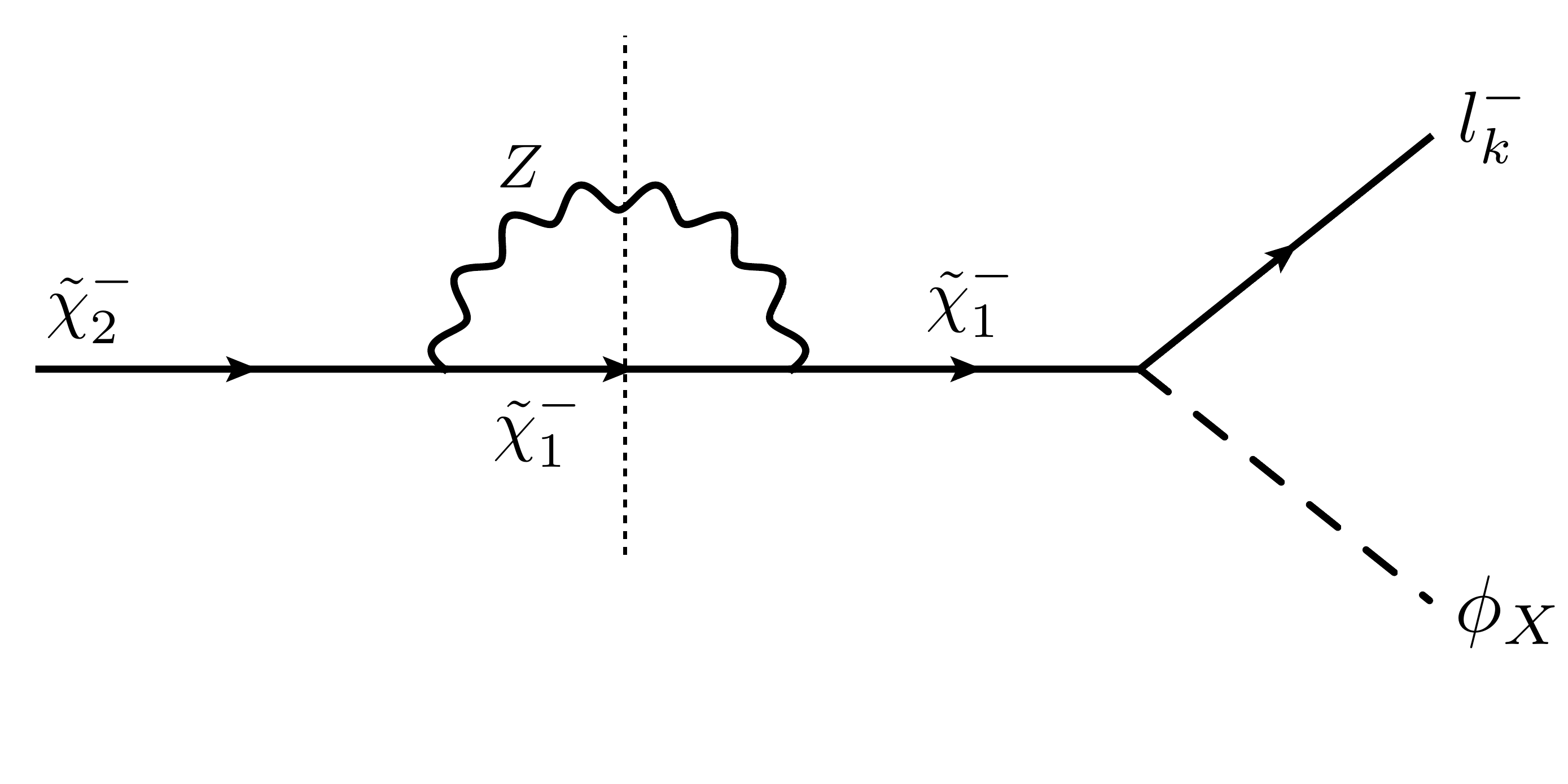}
\end{array}$ 
\caption{Examples of the diagrams that contribute to the asymmetry in the case of the
decays of the heavier chargino ${\tilde \chi}_2^-$ of the MSSM. Dashed vertical lines indicate position of the cuts.}
\label{treeplusloops} 
\end{figure}

In general a combination of intermediate MSSM states leads to asymmetries $\ep_a^-$ and $\ep_a^0$ for $a>1$.   
As an example, consider the decay of the heaviest chargino, $\tilde{\chi}^-_2$. The tree level decay  $\tilde{\chi}_2^-\rightarrow l_k^-\phi_X$ and the relevant loop diagrams that contribute to the generation of an asymmetry are depicted in Figure~\ref{treeplusloops}. (Additional diagrams involving the decay of the lighter chargino $\tilde{\chi}_1^-$ or the decay of the neutralinos will generally also contribute to the total asymmetry generated.)  We ignore diagrams with internal sleptons since they are proportional to lepton Yukawa couplings.  The necessary imaginary parts of these amplitudes arise from on-shell particles in the cut loops.  This simply means that for example in the first loop diagram only neutralinos appearing in the loop with masses satisfying $m_{\tilde{\chi}_i^0}< (m_{\tilde{\chi}_2^-}-m_W)$ contribute to the asymmetry.

We also require the decays to be CP-violating. The important phases for our example model are the phases in the chargino and neutralino sectors that feed into the mixing matrices, and therefore the diagrams in Figures~\ref{treeplusloops}, unsuppressed by small Yukawas.  The sizes of these phases are restricted by the measurements of various particle's EDMs \cite{rpp}\cite{edmstheory}, but phases up to $\sim \pi/30$ are allowed with only mild tuning.

\begin{figure}[t] 
\vspace{-15mm}
\bec
$\begin{array}{c}
\hspace{-6mm}
\includegraphics[width=0.35\textwidth]{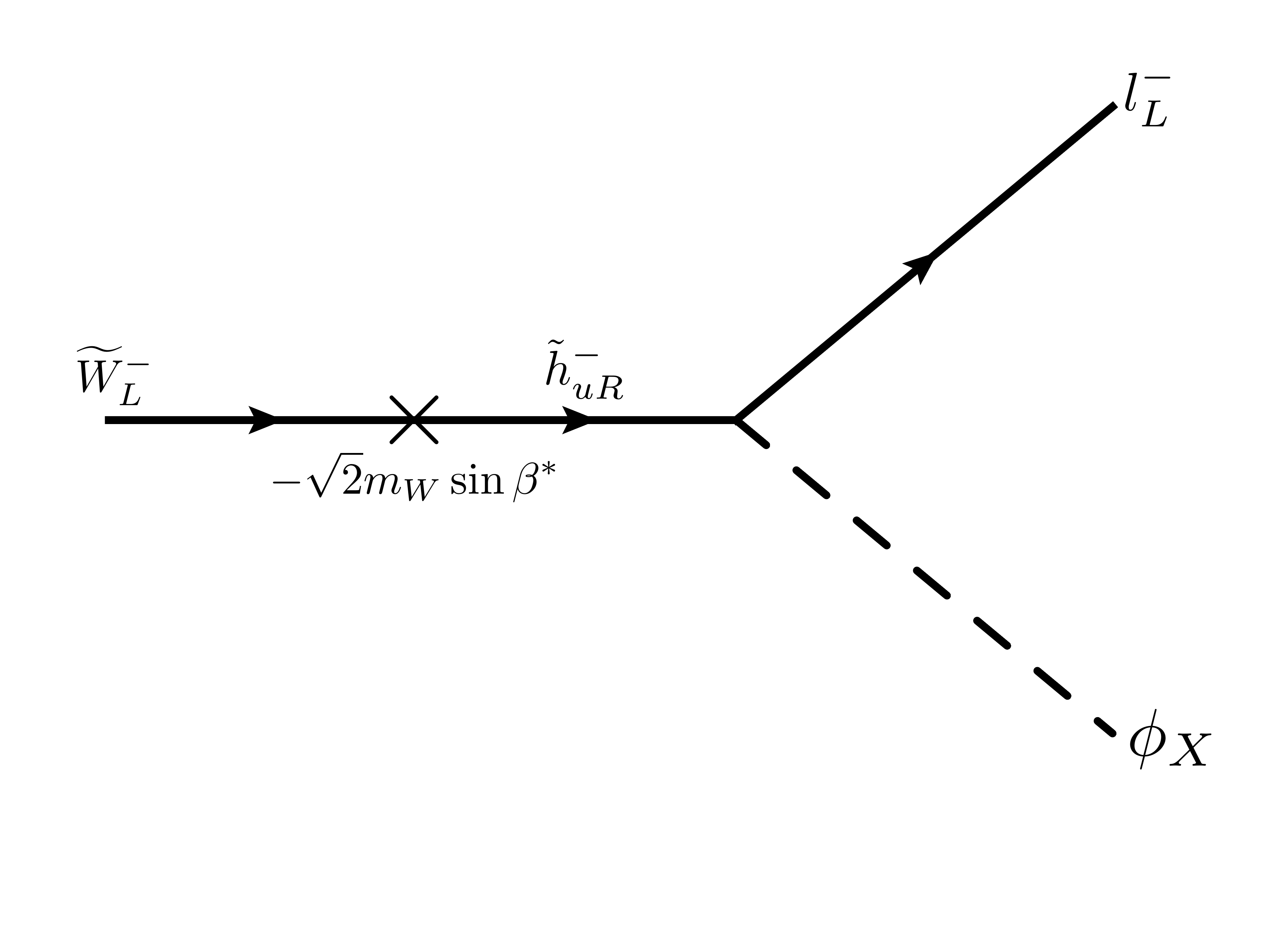}
\includegraphics[width=0.5\textwidth]{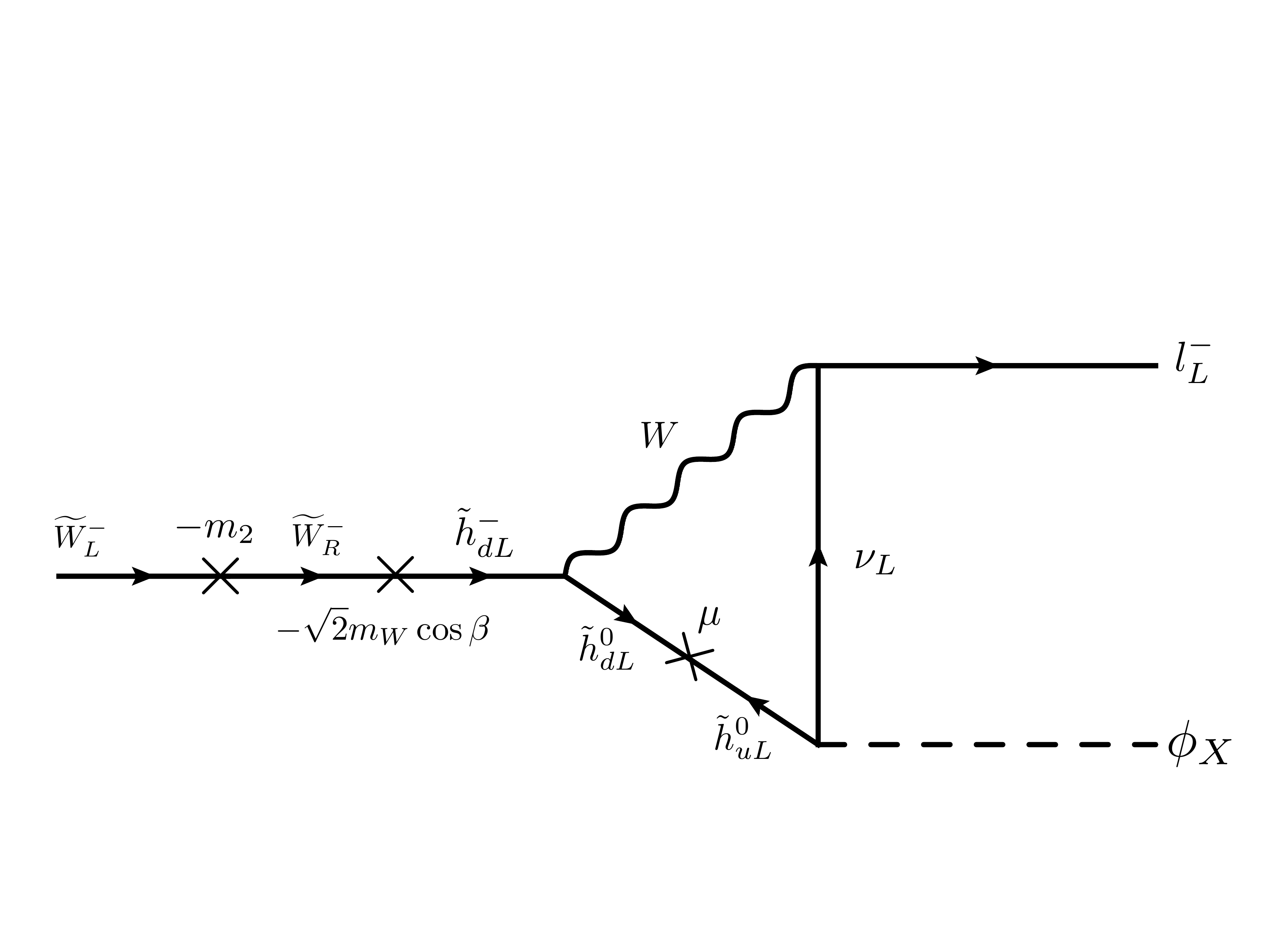} 
\end{array}$ 
\caption{Diagrams depicting the explicit mass insertions that encode the dependence of the asymmetry on the potentially CP-violating parameters in the gaugino-higgsino sector.} 
\label{massin} 
\eec
\end{figure} 

That the decays have physical CP-violation and lead to an asymmetry can be shown using the mass insertion approximation. Doing so enables us to construct the expression for the asymmetry in terms of the physical re-phase-invariant combinations of parameters in the gaugino-higgsino sector.  To demonstrate the mechanism we take a simplifying limit, $m_1 \rightarrow \infty$, such that we are left with three light neutralinos and two light charginos.  Initially the gaugino-higgsino sector contained two physical phases, after taking $m_1 \rightarrow \infty$ we are left with just one, namely $\phi_2=\arg{[\mu\; m_2\; b^*]}$ which can more conveniently be written as $\phi_2=\arg{[\mu\; m_2 \sin{2\be}]}$.  In addition we take
$m_2-\mu>m_Z$ with the absolute values of $m_2$ and $\mu$ being much larger that $m_Z$. In this limit we have a heavy neutralino and a heavy chargino (comprising of $\widetilde{W}^0$ and $\widetilde{W}^-$ respectively) both with approximate mass of $m_2$. We also have a further chargino with mass of approximately $\mu$ (comprised of the charged Higgsino states) and two neutralinos with masses also of approximately $\mu$ (comprised of the neutral Higgsino states).

We can now consider the decay $\tilde{\chi}^-\rightarrow l^-\phi_X$ in the weak eigenstate basis and by making chirality explicit we have $\widetilde{W}_L^-\rightarrow l^-_L\phi_X$. The tree level decay in the mass insertion approximation is depicted in Fig. \ref{massin}. The combination of coupling constants for this diagram is
\beq
C_{0}=-\sqrt{2}m_W\la^*\sin \be^*.
\eeq
One of the one loop diagrams that interferes with the tree graph is also sketched in Fig.\ref{massin}. The product of the one loop coupling constants for this diagram is
\bea\nonumber
C_{1}&=&-\frac{g^2}{\sqrt 2}m_2\cos\be\; m_W\mu\;\la^*.
\eea
Combining this with the tree-level coupling we find
\bea\nonumber
\Im{[C_{0}(C_{1})^*]}&=&\phantom{-}\frac{g^2}{2}m_W^2\abs{\la}^2\Im{[\mu^*m_2^*\sin 2\be^*]}
\eea
which together with eq.(\ref{cutformula}) shows that a CP-violating asymmetry is present in the decay of $\tilde{\chi}^-_2 \rightarrow \ell^- \phi_X$ when the loop amplitudes have on-shell cut diagrams with associated imaginary parts.  We can 
parameterise the resulting asymmetry as  
\bea
\ep =  \ep_0\al_w\sin \phi_2,
\eea
where $\al_w=g^2/4\pi$ and $\ep_0$ is a number that is determined by the neutralino and chargino mixing matrices and masses. The maximal size of $\ep_0$ is typically ${\cal O}(0.1)$, leading to values of $\ep\lsim10^{-3}$ if the unknown phase $\phi_2$ is ${\cal O}(1)$. 

We stress that the above simplifying limits are only taken so as to simply demonstrate the existence of a non-zero asymmetry in
heavier sparticle decays, and the limits are not a necessary part of our mechanism. For more realistic cases the exact size of the total asymmetry depends sensitively on the MSSM spectrum, via both the masses of the intermediate states that contribute and the details of the neutralino and chargino mixing matrices.

\subsection{Asymmetric freeze-in via operators of dimension 5}

Simultaneous DM and baryogenesis may occur via interactions that connect the MSSM sector to the
LSP $X$ chiral superfield at dimension 5, either via the superpotential
\beq
\De W \supset \frac{h_1}{M_*} {\bar U} {\bar D} {\bar D} X, ~~  \frac{h_2}{M_*} LQ{\bar D} X, ~~ \frac{h_3}{M_*}  LL{\bar E} X \label{eq:W-5}
\eeq
or via the Kahler potential
\beq
\De K \supset \frac{h_4}{M_*}L H_d^\dagger X, ~~  \frac{h_5}{M_*}L H_d^\dagger X^\dagger, ~~ \frac{h_6}{M_*}  L H_u X^\dagger   \label{eq:K-5}
\eeq
(if $h_1 \neq 0$ then the other couplings, including that of $LH_uX$, will be taken to be zero to forbid too fast proton decay).
Again asymmetric freeze-in takes place if the  hidden sector is initially at a different temperature to the visible sector and
the interaction strength is sufficiently small that the hidden sector does not thermalize with the visible sector.   Considering just the contribution from decays of non-LOSP superpartners, and assuming weak-scale superpartner masses, the correct asymmetry is produced for $M_*/h \sim 10^{(10,9)} (\ep / 10^{-3})^{1/2} \gev$ from operators of eqs.(\ref{eq:W-5}, \ref{eq:K-5}); intriguingly close to the intermediate scale.  

However, unlike the $LH_uX$
case these operators do not automatically lead to IR-dominated freeze-in of $\phi_X$.  In addition to the non-LOSP decay contribution, they lead to a UV-sensitive scattering contribution that grows linearly with the visible sector reheat temperature \cite{FIMP},  so that for large reheat temperatures the baryon and DM yields cannot be predicted.\footnote{Moreover, because of the potentially high temperatures at which the scattering contribution to the freeze-in yield occurs, the thermal corrections to masses and mixings (which we have hitherto been able to neglect at temperatures $T\sim m_w$) are non-negligible and have the potential to alter the asymmetries.}    The question is whether there is a range of reheat temperatures, $T_R$, not far above the superpartner masses,  where the IR dominated decay contribution dominates. 

For the operators of eq.(\ref{eq:W-5}), the leading non-LOSP decays to $\phi_X$ are three body.   These decays are phase space suppressed relative to the $2 \rightarrow 2$ scattering production of $\phi_X$ that takes place via the same diagram as for decays, but with a visible particle transferred from the final to the initial state.  This implies that the symmetric freeze-in is never dominated by decays, even if the $T_R$ is reduced to the mass of the superpartner whose decay is generating the freeze-in.  However, there may be a range of $T_R$ for which asymmetric freeze-in is dominated by the decays, because the rescattering is suppressed for the scattering contribution.  Consider the case of the $LLEX$ operator, with freeze-in contributions from the decays $\tilde{l} \rightarrow le \phi_X$ and $\tilde{l} \rightarrow l \tilde{\chi}$, and from the scatterings $\tilde{l} l \rightarrow e  \phi_X$ and $\tilde{l} l \rightarrow Z  \tilde{\chi}$.  The rescattering amplitude,  ${\cal A}_{12}$, occurs at tree level for decays, but is one-loop suppressed for scatterings, allowing DM and baryon asymmetries to be dominated by decays for $T_R < 10$ TeV.

Finally we consider the operators of eq.(\ref{eq:K-5}).
Inserting Higgs vevs into the last two interactions leads to kinetic mixing of the LSP, $\phi_X$, with a sneutrino.  Including soft supersymmetry breaking versions of these operators, the first gives $\tilde{\nu}/\phi_X$ mass mixing.  Hence, in all three cases, freeze-in can occur via 2 body decays:  $\tilde{\chi}^0 \rightarrow \nu \phi_X, \;  \tilde{\chi}^\pm \rightarrow l^\pm \phi_X, \; \tilde{\nu} \rightarrow (h,Z) \phi_X$ and $\tilde{l^\pm} \rightarrow (h^\pm,W^\pm) \phi_X$.   (For colored superpartners the decays are at least 3 body.)   Thus we expect a freeze-in abundance that is independent of $T_R$, for $T_R = 10$ TeV.  

For reheat temperatures above 10 TeV, asymmetric freeze-in from any operator of dimension 5 gives baryon and DM abundances that are linearly dependent on $T_R$; the $LH_uX$ operator is uniquely predictive for large $T_R$.   However, characteristic features of asymmetric freeze-in from dimension 5 operators, such as highly displaced baryon and/or lepton number LOSP decays at the LHC, can still be observed and tested.

\section{Prediction for the DM mass\label{massprediction}}

In the general case, where all or part of the asymmetry is generated before the electroweak phase transition (EWPT), the prediction for the $X$ mass depends upon the sphaleron-induced equilibration between $B$ and $L$.  In the non-supersymmetric case and assuming the asymmetry is fully generated before the EWPT, and the transition is not so strong that sphaleron processing immediately switches off below the EWPT temperature, the relation between $B$ and $B-L$ reads \cite{Harvey:1990qw,Inui:1993wv,Chung:2008gv} 
\beq
B =  \frac{8 N_G + 4 N_H }{22 N_G + 13 N_H }  (B - L)
\label{BLequilib}
\eeq
where $N_H$ and $N_G$ are the number of Higgs doublets and the number of particle generations respectively.
Eq.(\ref{BLequilib}) also assumes lepton flavour violation in equilibrium, and ignores potentially important plasma-mass corrections as we discuss below.

Since for all our inter-sector mediating operators $(B-L+X)$ is conserved, and eq.(\ref{BLequilib}) gives $B=c (B-L)$ for some spectrum
dependent factor $c$, in all cases where sphalerons are active one finds the relation $|B|=|c X|$ linking the baryon and $X$ number densities. Thus there arises a prediction for the DM mass depending on the visible-spectrum-dependent factor $c$, but independent of the details
of freeze-in.

For the default minimal SM with one Higgs doublet (and not the extended Higgs structure argued for in Section~\ref{nonsusy}) and a top quark mass not significantly above the lowest temperatures at which sphalerons are active one finds from eq.(\ref{BLequilib}), $B = (28/79) X$.  Thus  
\beq
m_X  = \frac{28}{79} m_p \frac{\Om_{d}}{\Om_b} \simeq  1.62\gev.
\label{Xmass1}
\eeq

Following the analysis presented in \cite{Inui:1993wv,Chung:2008gv} the size of $c$ varies with the details of the spectrum. For the default fully supersymmetric MSSM case, but with all superpartners heavy (such that eq.(\ref{BLequilib}) is still valid but the number of Higgs doublets is 2 instead of 1 as with the SM) we find $B= (8/23)(B-L)$ and so leads to a DM mass prediction
\beq
m_X  =  \frac{8}{23} m_p \frac{\Om_{d}}{\Om_b} \simeq  1.59\gev  ,
\label{Xmass2}
\eeq
pleasingly close to the SM value.

However, in the MSSM the variation in $X$ mass prediction due to the MSSM spectrum can be substantial \cite{Inui:1993wv,Chung:2008gv} (though we emphasize that once the MSSM spectrum is known there is a definite prediction for $m_X$).  Using expressions eq.(33) through eq.(40) of Ref.\cite{Chung:2008gv},
one finds that, in the situation where lepton flavour-changing interactions are operative, the largest factor is $B = 0.606 X$ implying
$m_X \simeq  2.76 \gev$.  This case primarily requires light LHD squarks.   On the other hand the smallest factor in the situation where lepton flavour-changing interactions are operative is $B = 0.079 X$ implying $m_X =0.36 \gev$.  This requires light RHD up squarks, LHD sleptons, and RHD selectrons.  In the case where lepton flavour violation is not operative then even more extreme values are conceivable depending
on the visible spectrum \cite{Chung:2008gv}, for example $B=-0.05X$ leading to $m_X=0.23 \gev$. 

It is possible that an even larger range occurs if one works at large $\tan \be$.   In this case there are plasma mass corrections, depending
on the thermal bath of real Higgs particles, enhanced by $1/( \cos\beta )^2$.   As long as the Higgs particles are not too heavy, the interaction with thermal bath of real Higgs particles can be large at large $\tan\be$ where the Yukawas for the leptons are large, even though $\vev{H_d}$ is small.\footnote{There may be other ways of altering the prediction for the $X$-mass.   For instance large individual lepton
numbers $L_i$, with total $L=\sum L_i$  almost cancelling to zero with associated $X$-number-density also suppressed, but because of plasma mass
effects in the MSSM sector a large $B$-number being produced from the large individual $L_i$'s as was attempted in a different context in Ref.\cite{MarchRussell:1999ig}.}.

Despite this dependence, it is important to emphasize that unambiguous predictions for the DM mass can be made once
the weak-scale visible spectrum is measured, and the form of the interactions is known.  Moreover, the equilibration
coefficient $c$ linking $B$ to $(B-L)$ is relatively insensitive to precise particle masses or sizes of couplings, depending
instead on qualitative features of the spectrum and interactions, allowing accurate predictions for the DM mass to be
made without similarly precise measurements of the TeV theory.

Finally, there exists one exceptional case:  If the dominant contribution to the $X$ asymmetry is generated below the EWPT and sphalerons are inactive, then any asymmetry in lepton number does not get transferred to baryon number, and only the ${\bar U} {\bar D} {\bar D} X$ interaction can give linked asymmetries in $B$ and $X$.  (Note that from Fig.~1 the dominant asymmetric
freeze-in yield from decays occurs at temperatures $\sim m_{NLOSP}/2$, requiring a light SUSY spectrum.  Since the
non-renomalizable ${\bar U} {\bar D} {\bar D} X$ interaction also has a $T_R$-dependent UV contribution, this
situation requires $T_R$ to be fine tuned to $\sim m_w$.)  Because in this case the $B$ asymmetry does not get
partially reprocessed into $L$, there is an unambiguous prediction for the DM mass:
\beq
m_X  =  m_p \frac{\Om_{d}}{\Om_b} \simeq  4.56\gev  .
\label{Xmass0}
\eeq

\section{Signatures \label{signatures}}

\subsection{A long-lived LOSPs at the LHC \label{tauLHC}}

Asymmetric Freeze-In requires the LOSP lifetime to be greater, usually much greater, than $10^{-13}$s to avoid thermalization of the two sectors, and can range up to the BBN bound, which is typically $10^2$s, and hence leads to events with displaced vertices or to decays of stopped LOSPs in the outer part of the detector.   In Section~\ref{model}, Asymmetric Freeze-In was found to be IR dominated if it arose via dimension 4 operators for any $T_R$, or via dimension 5 operators with $T_R \lsim 10 \tev$.   In these situations the LOSP lifetime is correlated to the DM/baryon abundance.  Here we explore the challenge of verifying the Asymmetric Freeze-In mechanism by measuring the LOSP lifetime.

The asymmetry in the yield of $X$ produced by freeze-in from visible sector particles of species $A$ is
\beq
\eta_X \;=\; C_{FI} M_{Pl} \sum_A \frac{\epsilon_A \Gamma_A \, g_A}{m_A^2}  \;=\; C_{FI} M_{Pl} \frac{\epsilon_a \Gamma_a \, g_a}{m_a^2} (1+f)
\label{etaX}
\eeq
where $\Gamma_A$ is the decay width of $A$ to final states containing $X$, $\epsilon_A$ is the CP violating fractional asymmetry in this width, and $g_A$ is the number of spin states of $A$.  The final expression in eq.(\ref{etaX}) follows from defining the largest contribution to $\eta_X$ as arising from $A=a$, so that $(1+f)$ can be viewed as a multiplicity factor taking account of other contributions.  We define $\Gamma_a/\Gamma_{\rm LOSP} = r \; m_a/m_{\rm LOSP}$.  In any particular model, if freeze-in is dominated by a single connector operator then $r$ will be independent of the coefficient of this operator, and depends only on visible sector parameters that can be measured.  The constraint that the $X$ asymmetry gives the observed dark matter abundance can then be translated into the decay length of the LOSP at the LHC
\beq
L \simeq 10 \; \mbox{m} \left(\frac{\ga}{2}\right) (1+f)r\left(\frac{\epsilon_a}{10^{-5}}\right) \left(\frac{m_X}{\gev}\right) \left(\frac{(10^2 \gev)^2}{m_{\rm LOSP} \, m_a}\right)  \left(\frac{g_a}{2} \right) \left(\frac{10^2}{g_*}\right)^{3/2} .
\label{decaylength}
\eeq
Here $\ga$ is the Lorentz boost factor, $g_*$ the effective number of degrees of freedom
at the epoch of freeze-in, and $g_a$ counts the spin states of $a$.  We have normalized the size of the CP-violating asymmetry to $\epsilon_a = 10^{-5}$, since this is midway in the allowed range of $10^{-(9-10)} \lsim \epsilon_a \lsim 10^{-(1-2)}$.   The upper bound arises because the asymmetry is a loop effect; the lower bound ensures that the symmetric freeze-in component does not bring the two sectors to a single temperature.  We stress that the range of $\epsilon_a$, and therefore of $L$, is very large.  Since the ATLAS and CMS detectors have size $\sim 10$ m, eq.(\ref{decaylength}) implies that in many cases a large fraction of all decays occur in the detector, allowing measurement of the LOSP lifetime.  However, the sensitivity of $L$ to $\epsilon_a$ and $r$, which could be much greater than unity, should be kept in mind.  

What are the prospects that eq.(\ref{decaylength}) can be verified? Both $m_a$ and $m_{\rm LOSP}$ can be measured by standard cascade decay techniques while, for a given SUSY spectrum, the $X$ mass is predicted via the procedure described in Section~\ref{massprediction}.   While the parameters $r$ and $f$ are model-dependent, they depend only on the spectrum of the visible superpartners.  Hence the biggest challenge is likely to be measuring the CP violating phase in $\epsilon_a$.  In the $LHX$ theory discussed in Section~\ref{LHX} we argued this arises from the phases in the Higgsino and gaugino mass matrices.   

\subsection{Baryon and/or lepton number violation at the LHC \label{BL-LHC}}

A striking, characteristic feature of Asymmetric Freeze-In is the occurrence of baryon and/or lepton number violation at the decay vertices of the LOSP.  It would be striking to directly observe baryon or lepton number violation at LHC.   We assume that the nature of the LOSP can be inferred from the cascade chains occurring at short distances.

Consider for definiteness the $LHX$ model with a charged slepton LOSP, such as the stau, as occurs in a sizeable portion of MSSM parameter space. Decays occur via the $A$-term associated with $\la  LHX$, or via $F$-term cross-interactions such as $\la \mu^* \, 
h_d^* {\tilde l}  \phi_X$.  In either case the stau, identified as a lepton number carrying state by observation of the prompt leptonic part of the decay chain, 
decays to a virtual charged Higgs and missing energy.  
Kinematics will show that the decay is two body, and therefore that the missing energy must be carried by a scalar, not a neutrino.  In this example it would be clear that lepton number violation had occurred -- the leptonic nature of the LOSP would be established and the final state is the two-body mode $h^\pm \phi_X$.  

Lepton number violation could be similarly established much more generally.  For any of the operators $L H_u (X, X^\dagger), L H_d^\dagger (X, X^\dagger)$, the violation of lepton number would be apparent in all the leading two-body decay modes, no matter what the nature of the LOSP
\beq
\tilde{l}^\pm;\tilde{\nu} \rightarrow (h^\pm, W^\pm; h,Z) \phi_X
\eeq
\beq
\tilde{\chi}^\pm;\tilde{\chi}^0 \rightarrow (l^\pm; \nu) \phi_X
\eeq
\beq
\tilde{q} \rightarrow j l^\pm \phi_X \hspace{1in} \tilde{g} \rightarrow jj l^\pm \phi_X
\eeq
where $j$ is a jet.  The exception is the neutralino LOSP, since $\tilde{\chi}^0 \rightarrow  \nu \phi_X$ is invisible.  In this case, assuming a short enough lifetime, lepton number violation could be established via the subdominant three-body mode $\tilde{\chi}^0 \rightarrow l^\pm W^\mp \phi_X$.   For the operator $LL\bar{E}X$, the dominant LOSP decays involve 2 or 3 leptons in addition to the LSP $\phi_X$.  In cases where the missing energy is shared between the LSP and a neutrino, establishing lepton number violation will be difficult; but it should be possible in decays such as $\tilde{\chi}^+ \rightarrow l^+ l^+ l^- \phi_X$.

What are the prospects of discovering baryon number violation at the LHC from the decay of the LOSP via the $\bar{U}\bar{D}\bar{D}X$ interaction?  We again assume that the nature of the LOSP has been determined from short distance cascades, and that the spin of the LOSP is known either by direct measurement or by its supersymmetric interpretation.  A color neutral neutralino or chargino LOSP will decay to three jets $(\tilde{\chi}^\pm,\tilde{\chi}^0) \rightarrow jjj$.  A statistical study of these events would show that all jets originate from quarks and anti-quarks rather than gluons; furthermore, the identification of the sources of the jets as $qqq$ rather than $\bar{q}q g$ would be required by angular momentum conservation, establishing baryon number violation.   A dominant decay mode of any slepton LOSP would be to three jets and a charged lepton, $(\tilde{l}^\pm,\tilde{\nu}) \rightarrow l^\pm jjj$.  Similar reasoning would require a $qqq$ rather than a $\bar{q}qg$ origin of $jjj$.   Finally, consider a squark LOSP, such as $\tilde{t}$.  It would hadronize and stop in the detector as a $R$ hadron.  However, the spectator quarks are irrelevant to the decay, which would amount to $\tilde{q} \rightarrow jj$.  Statistically the jets would be identified as originating from a high energy quark and, furthermore, interpreting one as a gluon jet would violate angular momentum, and interpreting both as gluons would violate color.

\begin{figure}
\vspace{-2cm}
\centerline{\includegraphics[width=12cm]{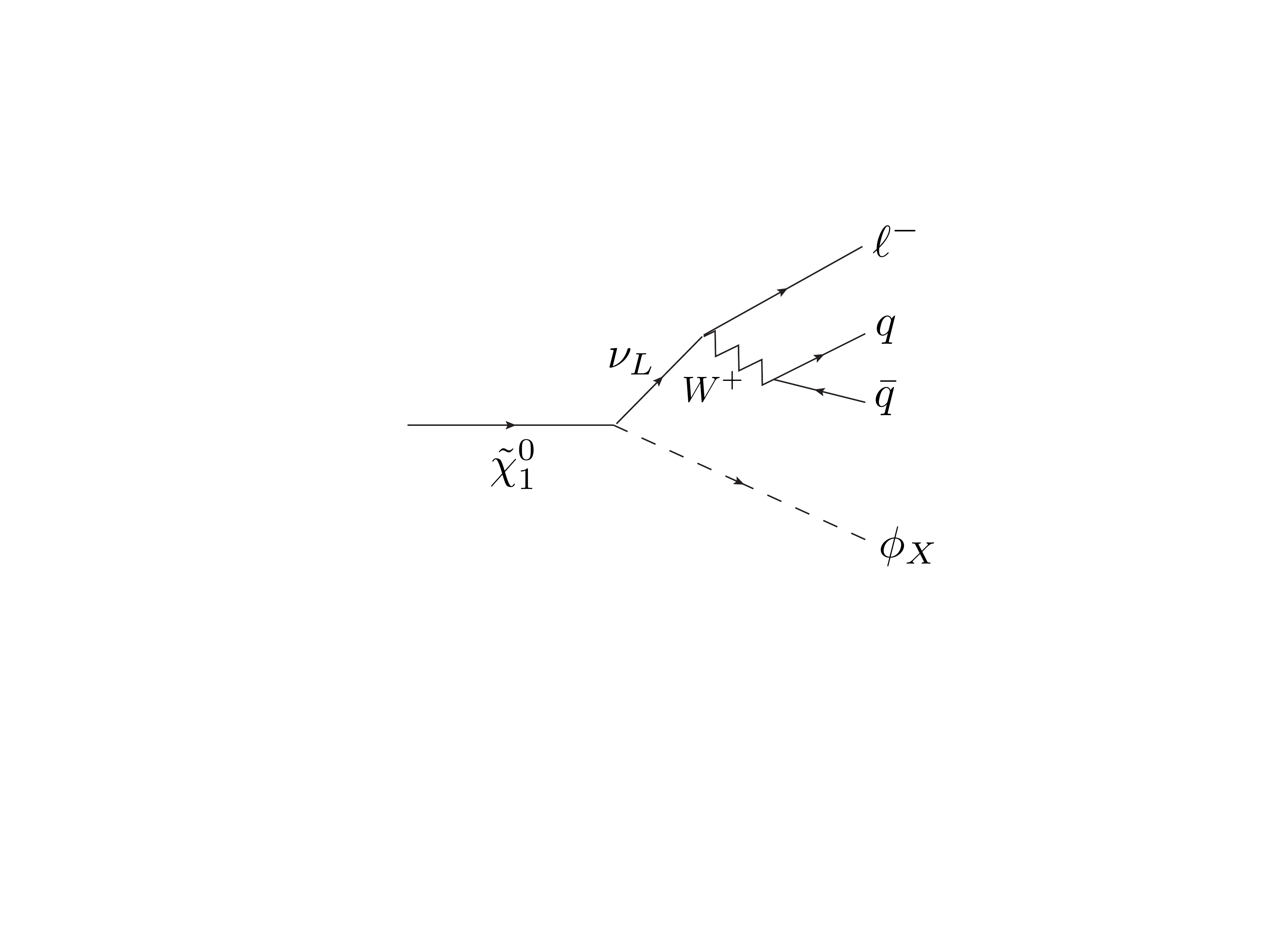}}
\vspace{-2.5cm}
\caption{Sub-leading contribution to neutralino LOSP decay allowing measurement of inter-sector coupling $\la$ in the case
of $\la LHX$ interactions.}
\label{neutralinodecay2}
\end{figure}


\subsection{EDMs and CP violation at a linear collider}

Once the low-lying supersymmetric spectrum is known and the inter-sector coupling $\la$ and the DM mass $m_X$ are measured (via, respectively, measurements of the LOSP lifetime, and dedicated studies at a precision collider experiment), then the master formulae
eqns.(\ref{eq:etaX}) and (\ref{FIyield}) lead to a prediction for the CP-violating asymmetry factor $\ep$.  Since we assume the low-energy SUSY spectrum to be known, this in turn leads to a prediction for the CP phases, eg, $\phi_1$ and $\phi_2$, and therefore, baring fine-tuned
cancellations, to predictions for CP violating EDM's and collider observables which can further confirm the asymmetric FIMP mechanism.
Even if the first two generations of squarks and sleptons are superheavy, thus ameliorating the SUSY CP problem there are, at two loops,
irreducible contributions to EDMs arising from chargino and neutralino loops which are accessible to current and future EDM experiments.
A linear collider running on the neutralino or chargino resonances also has the capability to measure the gaugino-higgsino sector CP-violating angles $\phi_{1,2}$  if they are not too small \cite{ILC}, and thus completely confirm the asymmetric freeze-in mechanism.

\subsection{Signatures from the $X$ sector \label{Xsector}}

As discussed in the introduction, a crucial part of our theory of matter genesis is that the symmetric part of the DM density be efficiently annihilated away, leaving just the irreducible asymmetric contribution.  This can lead to distinctive signatures in its own right.  

For example, consider adding to the SUSY model outlined in Section~\ref{model} the superpotential terms  
\beq
W_{\rm X-sector} =  \la' X X^c Y  +  m_X  X X^c + \la^{''} Y H_u H_d + m_Y Y^2
\label{Xsector1}
\eeq
where $X^c, Y$ are SM-singlet chiral superfields with, respectively, $R_p=-1$, $Q_X =-1$ and $R_p=+1$, $Q_X =0$,  and $\la^{''}\sim \la \ll  \la'$.
Corresponding soft terms, including $({\rm mass})^2$, $A$ and $B$ terms, are also present.  If the intra-sector coupling $\la'$ is not very small, and the scalar masses satisfy $m_{\phi_Y} < m_{\phi_X}$, then efficient annihilation of the symmetric part of the $X$-density to $\phi_Y$ states quickly occurs, these states later decaying to SM degrees of freedom via the $\la^{''}$ coupling.\footnote{Note that because of the limits on elastic self-interactions among DM particles \cite{SIDMlimits} the coupling $\la'$ cannot be too large for light $m_{\phi_Y}\ll \gev$.  However, due to the efficiency of matter-anti-matter annihilation in the presence of an asymmetry this bound is easily satisfied.}

Alternatively, the annihilation of the symmetric part may be to exotic light states. For example consider a model with states charged under a hidden $U(1)'$ gauge group with hidden photon and photino $(\ga', {\tilde \ga}')$ and gauge coupling $g'$ and matter superpotential
\beq 
W_{\rm X-sector} = \la' XYV + m_X  X X^c + m_Y  Y Y^c + m_V  V V^c
\label{Xsector2}
\eeq
where the $U(1)_X$ and $U(1)' $ charges are respectively $(1,0)$, $(-1,1)$ and $(0,-1)$ for $X,Y,V$  (and opposite for $X^c,Y^c,V^c$). 
In this case the intra-sector couplings $\la',g'$ allow the symmetric part to annihilate away to the hidden photon
via intermediate $Y$ and $Z$ states.   Note that this does not imply the existence of a new long-range force acting on DM, as, unlike
the global $U(1)_X$, the $U(1)'$ gauge symmetry can be spontaneously broken.    As long as $m_{\ga'}/\ep \ll m_{\phi_X}$ the asymmetric
$\phi_X$ density dominates the DM density independent of the hidden-sector freeze-out dynamics of $\ga'$.  However the requirement for
$m_{\ga'}$ to be small introduces another mass scale that requires explanation.  
 
Other interesting phenomenology concerns
the lightest $R_p$-even state that carries the $U(1)_X$ quantum number, and thus, effectively, $B$ or $L$ number depending on the mediating operator, as this state must late decay back to SM states, independent of the details of the $X$ sector.  Consider, for example, the case where the relevant state is the fermion $\psi_X$ itself (or in more complicated hidden-sector models a fermion $\eta$ containing some admixture of $\psi_X$).

Both the symmetric and asymmetric $\psi_X$ densities are set by model-dependent freeze-out dynamics in the
hidden sector.   This dynamics must be efficient enough to remove the initial $Y_X$ symmetric density
$n_{\phi_X} + n_{\phi_X^*}$ of $\phi_X$ down to levels $< 0.1 \eta_X = 0.1 \ep Y_X$ if we are to have a reasonably precise linking of
DM to baryon density.   The same freeze-out dynamics also determines how efficiently the symmetric and asymmetric density
in heavier states carrying $U(1)_X$ number is transferred to the lightest such state, which we assume to be $\phi_X$.
In particular, if $m_{\psi_X} > m_{\phi_X}$, then, via the $t$-channel exchange of $Q_X=0$ $R_p$-odd fermion states, there exists the
scattering process $\psi_X \psi_X \to \phi_X \phi_X$ reducing the number density of $\psi_X$ and moving the asymmetry to $\phi_X$.
At some temperature this process freezes-out leaving a relic density of $\psi_X$, conceivably as much as $0.1 \eta_X$, which then can late decay to SM states.

The situation with the $X \bar{UDD}$ mediating operator is the most interesting.  The neutral $R_p$-even $\psi_X$ fermion state effectively carries baryon number, so cannot be lighter than the proton, or unacceptably fast nucleon decay results.   Instead there typically exist open channels such as $\psi_X\to n\pi^0$ or $\psi_X \to p \pi^-$ with associated lifetime estimated to be 
\beq
\tau_{\psi_X} \simeq  10^{4}  \left( \frac{ 0.2 }{ \al_s } \right)^4 \left(\frac{M_*}{10^{9}\gev}\right)^2 \left( \frac{M_{susy} }{ 500\gev } \right)^2 \left( \frac{0.1}{ C } \right)^2 \left( \frac{2 \gev }{ m_{\psi_x} } \right)^5  sec
\eeq  
where $C$ is a hadronic matrix element, and $M_{susy}$ is an effective average squark and gluino mass.  Thus the state $\psi_X$ can be quite long lived with ${\cal O}(1)$ branching ratio to nucleons, giving rise to potentially important modifications of BBN predictions via late injection of energetic hadronic final states depending on the hidden-sector freeze-out density of $\psi_X$ states \cite{Jedamzik:2006xz}.  \footnote{Note that for lifetimes $\tau_{\psi_X}\gsim 10^4 s$ the upper limits on the energetic nucleon injection due to decays are quite stringent $\Om_{\psi_X} h^2 B_h\lsim 10^{-4}$.   We thank Karsten Jedamzik for earlier discussions on this issue.  We will return to the model-dependent phenomenology associated with the $X$-sector in a later publication \cite{upcomingwithKJ}.}

Consider, on the other hand, the $\la X L H_u$ case: Now $\psi_X$ decays via virtual $W,Z$ with lifetime 
\beq
\tau_{\psi_X} \simeq  10 \left( \frac{10^{-10}}{\la} \right)^2  \left( \frac{2 \gev }{ m_{\psi_x} } \right)^3  sec
\eeq  
to partonic final states $\bar\nu f {\bar f}$, or $\ell^+ \bar{f_u} f_d$.
Since these final states, for the $\psi_X$ masses being considered, imply a very small branching ratio to nucleons, the effect on BBN abundances is small \cite{Jedamzik:2006xz}.   For the other possible mediating operators where $\psi_X$ effectively carries lepton number similar results apply, although BBN signals can arise in limited parts of parameter space.

\begin{figure}
\vspace{-2cm}
\centerline{\includegraphics[width=12cm]{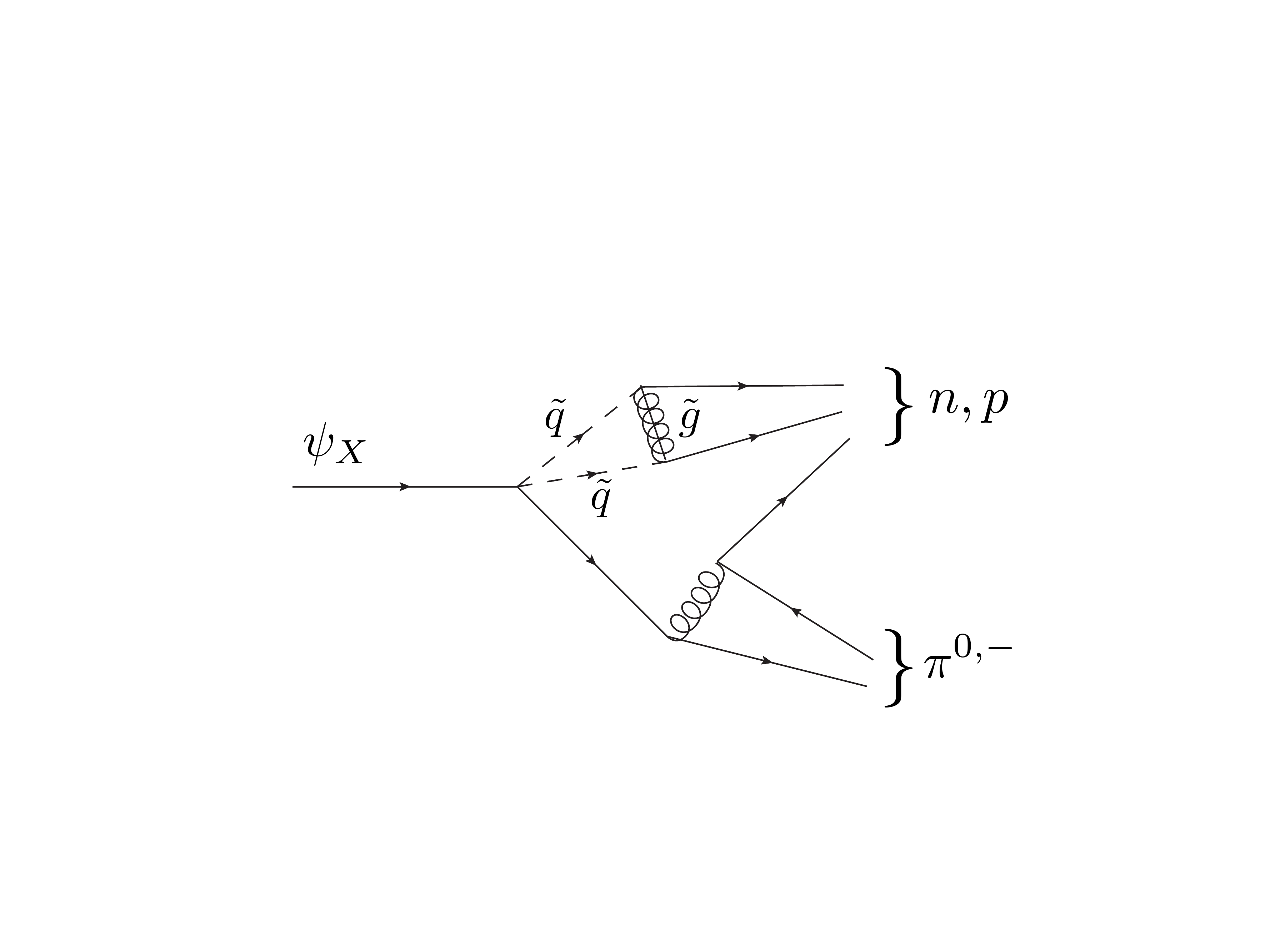}}
\vspace{-1.5cm}
\caption{Contribution to the decay of lightest $X$-sector, $R_p=+1$ state carrying $U(1)_X$ and thus baryon number in the case
of the $X {\bar U D D}$ inter-sector interaction.}
\label{psiXdecay}
\end{figure}

\section{Comments \& Variations \label{variations}}

\subsection{Non-supersymmetric models\label{nonsusy}}

Non-supersymmetric models implementing our mechanism are also possible.  As explained in Section~\ref{intro} in addition to 
(an almost exactly) conserved $U(1)_{(B-L-X)}$-number, a second DM-stabilizing symmetry, such as $R$-parity, is required. 
We thus complement the SM with a new sector, and impose a discrete symmetry, most easily a $Z_2$, under which both some SM and
$X$-sector states transform, with the lightest $Z_2$-odd state being an $X$-sector state also carrying non-zero $U(1)_{(B-L+X)}$-number.
A new possibility, compared to the supersymmetric case with $R$-parity as the stabilizing symmetry, is that the stable DM state
can be a fermion $\psi_X$.   For example, consider enhancing the SM with an $X$-sector, a second scalar Higgs SU(2)-doublet state, $H_2$, and a number of singlet scalars $S_i$ ($i=1,...K$) with Lagrangian
\beq
\De{\mathcal L} =  \la L H_2 \psi_X + {\rm quartic~and~quadratic~scalar~terms} + X{\rm sector}
\eeq
where $L$ is the usual leptonic fermion doublet, $H_1$ is the SM Higgs, and $\psi_X$, $H_2$, and at least one of the $S_i$ are $Z_2$-odd, with all other fields being $Z_2$ even.  The masses and couplings must be such that none of the $Z_2$ scalars acquires a vacuum expectation value and breaks the stabilizing symmetry.   A counting argument shows that for $K$ large enough there exist physical, irremovable phases in the complex couplings, and thus CP-violating decays of the heavier $S_i$ states to $\psi_X \nu_L$ via the portal interaction $\la L H_2 \psi_X$ can occur.   Since the DM particle is now a (SM-neutral) Dirac fermion, a small dipole moment interaction with the SM is now allowed, unlike the case of scalar DM.  

\subsection{Origin of the DM mass}

A full explanation of the baryon-to-DM energy density ratio requires more than just linked number densities for the DM and baryons: One requires a motivation for the DM mass to lie roughly near the proton mass in line with the values for $m_X$ given
in Section~\ref{massprediction}.  Such an explanation requires going beyond the IR effective theory that we have focussed upon so far.   

Our setup is most naturally realized in the context of the multiple sequestered hidden sectors expected in string or extra-dimensional UV completions of the SM, as discussed recently in \cite{hiddensectors,hiddensectors2}.  
Consider the supersymmetric case, where the DM FIMP is the true LSP, the scalar component of the R-parity odd $X$ field. As argued in \cite{hiddensectors2}, the large number, ${\cal O}(10^2)$, of hidden sectors expected in a typical supersymmetric string compactification implies that it is unlikely for the true LSP to be in our sector.  If SUSY is broken at high scales and the breaking is mediated by supergravity then the scales of soft terms in the various sectors is distributed around the weak scale, but with variations arising from the coefficients of the higher-dimension Kahler and superpotential operators involving the SUSY-breaking spurion.   If anomaly mediation of SUSY-breaking dominates, all sectors gain soft-masses proportional to the beta-functions and anomalous dimensions in the sector in question, and so are quadratically sensitive to the variations of the ${\cal O}(g)$ couplings in each sector.  In either case a range of hidden sector LSP masses 
below the weak scale arises independent of the small coupling between the sectors.

\subsection{Relic LOSP decay}

In many cases a significant abundance of LOSPs will freeze-out in the visible sector.  For example, if the LOSP is a weakly interacting neutralino the abundance left after freeze-out can naturally be larger than $\Om h^2 =0.1$. These LOSPs are, of course, not stable and will decay via the feeble coupling into the hidden-sector.   Because the resulting matter density in the hidden sector is reduced relative to the original LOSP density by a factor $m_X/m_{LOSP} \sim 10^{-2}$ this extra contribution to the hidden sector density typically cannot significantly change the final DM density.  Nevertheless in some special cases \cite{hiddensectors2} the LOSP freeze-out density is large enough that this decay can lead to a significant increase in the abundance of $\phi_X$ and $\phi_X^*$ particles (the decays do not generate or erase the asymmetry, equal numbers of $\phi_X$ and $\phi_X^*$ particle will be produced).  If this process happens after the symmetric component of the DM density has annihilated away and dropped out of thermal equilibrium (i.e. the $\phi_X$ DM has completed freeze-out in the hidden sector) we will have an additional symmetric contribution to the DM relic abundance and so we loose the ability to explain the ratio $\Om_{d}/\Om_b$ with linked asymmetries.  We now show that this is not the case.

If the temperature at which the LOSPs decay is high enough in the $X$-sector that the $\phi_X$ and $\phi_X^*$ particles are still in thermal equilibrium then the extra abundance is simply annihilated away with the rest of the symmetric abundance of $\phi_X$ generated through the freeze-in.  The energy radiated into the hidden sector during freeze-in heats the hidden sector to a temperature $T_{X,i}$ at the end of freeze-in. Assuming the initial hidden-sector temperature is zero, the relationship between the temperatures in the visible and hidden sectors just after the completion of freeze-in is
\beq
\frac{T_{X,i}}{T_{\rm{vis}}}\simeq 0.05  \left(\frac{10^{-3}}{\ep}\right)^{1/4}\;\left(\frac{10}{g_X}\right)^{1/4}\;\left(\frac{g_*}{100}\right)^{1/4}
\label{hstemp}
\eeq
where $g_X$ is the number of relativistic degrees of freedom in the hidden sector. 
On the other hand, the temperature in the visible sector at the epoch of LOSP decay is
\beq
T_{\rm{decay}}^{\rm{vis}}=
15.5 \gev \left(\frac{m_{\rm{LOSP}}}{100 \gev}\right)^{1/2}\;\left(\frac{\la}{10^{-8}}\right)\;\left(\frac{g_*}{100}\right)^{1/4} .
\eeq
Using this and eq.(\ref{hstemp}), and ignoring, for simplicity, any changes in the number of relativistic hidden-sector degrees of freedom, we find the temperature in the hidden-sector at which the LOSPs decay to be
\beq
T_{X\rm{LOSP - decay}}\simeq 770\mev \left(\frac{m_{\rm{LOSP}}}{100 \gev}\right)^{1/2}\;\left(\frac{\la}{10^{-8}}\right)\;\left(\frac{10^{-3}}{\ep}\right)^{1/4}\;\left(\frac{g_X}{10}\right)^{1/4}.
\label{txlosp}
\eeq
This is to be compared to the typical value for the freeze-out temperature, $T_{X,fo}$, in the hidden sector:  Given a hidden sector mass scale $\sim \gev$ this will be $T_{X,fo} \sim few 10\mev$, where we have allowed for a wide variation in the size of the intra-hidden-sector couplings leading to freeze-out, such as $\la'$ in eq.(\ref{Xsector1}).

Thus for typical values of the parameters involved the LOSPs decay sufficiently early such that hidden-sector freeze-out in still in operation, and the extra symmetric DM density is annihilated away independent of the LOSP visible-sector freeze-out density.

\section{Conclusions \label{conclusions}}

We proposed a unified theory of dark matter genesis and baryogenesis utilizing the thermal freeze-in mechanism of DM
production involving decays from visible to hidden sectors.   Calculable, linked, asymmetries in baryon number and DM number are produced by the
interaction mediating between the two sectors together with CP-violating phases in the MSSM (or extended Higgs sector of the SM).
The out-of-equilibrium condition necessary for baryogenesis is provided by the different temperatures of the visible and hidden sectors. 
Our theory explains the observed coincidence between the DM and baryon densities, and specific realizations of our mechanism
can be completely tested by a combination of collider experiments and precision tests.  Characteristic signals of this mechanism
are spectacular, including long-lived metastable states late decaying via apparent baryon-number or lepton-number
violating processes at the LHC.  These features are mandated in our mechanism, being directly related to the necessarily
feeble inter-sector interaction that connects $(B-L)$-number to hidden sector $X$-number.
Other signals include EDMs correlated with the observed decay lifetimes and masses and within reach of
planned experiments.  Depending on the details of the necessary hidden sector interactions, late decays of additional
$X$-sector states back to the visible sector are possible, leading to potential alteration of big
bang nucleosynthesis predictions.

\section*{Acknowledgements}
We gratefully thank Karsten Jedamzik for earlier collaboration on the freeze-in mechanism and FIMPs.  LH and SMW would like to acknowledge the Dalitz Institute for Fundamental Physics and the Department of Theoretical Physics, Oxford University for hospitality during portions of this work, while JMR  would like to thank the particle theory groups at both UC Berkeley and Stanford University for hospitality during the inception of this work.   JMR also acknowledges support by the EU Marie Curie Network ÒUniverseNetÓ (HPRN-CT-2006-035863), and by a Royal Society Wolfson Merit Award. SMW also thanks the Higher Education Funding Council for England and the Science and Technology Facilities Council for financial support under the SEPNet Initiative.  The work of L.H. was supported in part by the Director, Office of Science, Office of High Energy and Nuclear Physics, of the US Department of Energy under Contract DE-AC02-05CH11231 and by the National Science Foundation on grant PHY-0457315.

\bibliographystyle{JHEP}

\end{document}